\documentclass[twocolumn,english,aps,pra,longbibliography,superscriptaddress,amsmath,amssymb,floatfix]{revtex4-2}
\usepackage[utf8]{inputenc}
\usepackage{microtype}
\usepackage{graphicx}
\usepackage{amsmath}
\usepackage{amssymb}
\usepackage{mathtools}
\usepackage[colorlinks=true,urlcolor=blue,citecolor=blue,linkcolor=blue]{hyperref}
\usepackage{natbib}
\usepackage{physics}
\usepackage{cleveref}
\usepackage{sidecap}
\usepackage{makecell}
\usepackage[linesnumbered,lined,commentsnumbered]{algorithm2e}
\Crefname{algocf}{Algorithm}{Algorithms}
\usepackage{layouts}
\usepackage{dcolumn}
\usepackage{bm}
\usepackage[percent]{overpic}
\usepackage{dsfont}
\usepackage{comment}
\usepackage{float}
\usepackage{minitoc}
\usepackage{titletoc}
\usepackage{pifont}     
\usepackage{xcolor}
\usepackage{soul}
\usepackage{tabularx}
\usepackage{booktabs}
\usepackage{amssymb}
\usepackage{minitoc}
\usepackage{hyperref}

\makeatletter

\newcommand{\appendixtableofcontents}{%
  \@starttoc{apx}
}

\makeatother

\begin{document}
\preprint{APS/123-QED}

\title{Enhancing the phase sensitivity of a Mach--Zehnder interferometer beyond the Heisenberg limit through dynamic squeezing}

\date{\today}

\author{Lakshya Bhardwaj}
\thanks{Both authors contributed equally to this work.}
\affiliation{\textit{Department of Physics, Arizona State University, Tempe, Arizona 85287, USA}}
\affiliation{\textit{School of Electrical, Computer, and Energy Engineering, Arizona State University, Tempe, Arizona 85287, USA}}
\author{Ankit Tiwari}
\thanks{Both authors contributed equally to this work.}
\affiliation{\textit{School of Electrical, Computer, and Energy Engineering, Arizona State University, Tempe, Arizona 85287, USA}}
\author{Mohan Sarovar}
\affiliation{\textit{Quantum Algorithms and Applications Collaboratory, Sandia National Laboratories, Livermore, California 94550, USA}}
\author{Christian Arenz}
\affiliation{\textit{School of Electrical, Computer, and Energy Engineering, Arizona State University, Tempe, Arizona 85287, USA}}

\begin{abstract}
We propose a method to enhance the phase sensitivity of a Mach-–Zehnder interferometer, independent of its input states. This is achieved by applying squeezing sequences to one arm of the interferometer. By alternating between squeezing along orthogonal quadratures, we demonstrate that the phase sensitivity of a Mach--Zehnder interferometer can be generically enhanced. Since this enhancement is independent of the input states, the proposed dynamic squeezing method allows for further improvement in phase sensitivity when combined with non-classical input states. We compare the dynamic squeezing approach with existing quantum sensing protocols and show that super-Heisenberg scaling can be achieved. Finally, we demonstrate that the scheme is robust against moderate photon loss.
\end{abstract}

\maketitle

\section{Introduction}

Quantum sensing leverages quantum effects to infer desired physical quantities more precisely \cite{RevModPhys.89.035002,pirandola2018advances}. Often such quantities, including magnetic and electric field amplitudes \cite{budker2007optical}, gravitational waves \cite{abbott2016observation} etc., manifest as a phase shift on a probe field \cite{helstrom1969quantum,paris2009quantum, giovannetti2006quantum,giovannetti2011advances}. The paradigmatic approach to estimating phase shifts is the Mach--Zehnder interferometer (MZI) \cite{scully1997quantum,DEMKOWICZDOBRZANSKI2015345}, see Fig.~\ref{fig: MZIschematic}(a). 

When the input into each arm of a MZI is a coherent state of light, the phase uncertainty of the interferometer is limited by the standard quantum limit (SQL), $\Delta\phi \sim 1/\sqrt{\mathcal{I}}$, where $\mathcal{I}$ is a measure of the intensity of the input fields \cite{giovannetti2004quantum,giovannetti2006quantum,DEMKOWICZDOBRZANSKI2015345}. There exist two different ways for improving the precision beyond the SQL using quantum effects. In the first strategy, the SQL is surpassed by using non-classical input states \cite{dowling2008quantum, PhysRevLett.85.2733, joo2011quantum,joo2012quantum, gerry2010heisenberg,liu2016quantum,PhysRevA.90.025802, campos2003optical,tan2014enhanced,birrittella2012multiphoton, bondurant1984squeezed}, such as squeezed states and entangled states, to achieve \emph{Heisenberg scaling} with input light intensity and saturate the Heisenberg limit (HL), $\Delta\phi \sim 1/\mathcal{I}$ \cite{PhysRevLett.71.1355,PhysRevLett.77.2352,giovannetti2006quantum}.  It is common to refer to any reduction in phase uncertainty beyond the SQL as improved \emph{phase sensitivity}, with the notion of sensitivity being the inverse of parameter uncertainty. In the second strategy, a non-Gaussian dynamical process is used that transforms the quantum state of light inside the interferometer to improve the phase sensitivity \cite{LUIS20048, PhysRevLett.98.090401, boixo2008quantum, napolitano2011interaction, ddboixo2008quantum, Woolley_2008,PhysRevA.65.025802,PhysRevA.80.032103,PhysRevLett.100.220501}. Such schemes, which require high-order non-linearities such as a self-Kerr interaction, can even surpass the HL, and are only limited by the degree of non-linearity.  Thus, even without entanglement in the input states, merely modifying the system dynamics can substantially enhance the phase sensitivity towards achieving super-Heisenberg scaling.  Another example is the SU(1,1) interferometer \cite{PhysRevA.33.4033, hudelist2014quantum, Chekhova:16}, where beamsplitters are replaced by optical parametric amplifiers. Nevertheless, these modifications of MZI dynamics have proved to be impractical; sufficiently strong high-order non-linear interactions are difficult to realize in low-loss media, and although coherent-boosted SU(1,1) interferometers \cite{plick2010coherent} have been experimentally realized, they fall short of reaching the HL \cite{hudelist2014quantum}.

In this work, we develop a modified MZI protocol that uses purely Gaussian operations within the MZI loop to dramatically reduce the estimated phase uncertainty, even beyond the HL. Our protocol utilizes a sequence of squeezing operations in one arm of the interferometer that alternates between squeezing along orthogonal quadratures; a scheme referred to as \textit{Hamiltonian amplification} (HA) \cite{arenz2020amplification, Tiwari:25,tiwari2026amplification}, which has been experimentally demonstrated in a trapped ion system \cite{burd2024experimental}. The modification of the dynamics we utilize only requires single-mode squeezing, which can be achieved with lower-order non-linearities, \emph{e.g.,} phase sensitive parametric amplification using a $\chi^{(2)}$ non-linearity. Crucially, the phase sensitivity enhancement arises here entirely from the modification of the dynamics, requiring neither entanglement nor squeezing in the initial state. This suggests the potential to combine this protocol with a non-classical input state for even greater sensitivity. Indeed, we show that in this way super-Heisenberg scaling can be obtained.

The manuscript is organized as follows. In Sec.~\ref{sec: Enhancement of MZI Sensitivity via Hamiltonian Amplification}, we start by introducing a conventional MZI setup with a photodetector at one output port --- referred to as the single intensity detection scheme. We then describe the HA protocol and show that the phase sensitivity of a MZI can be enhanced through HA regardless of the state considered at the input. In Sec.~\ref{subsec:Phase sensitivity with coherent state of light as input}, \ref{subsec:Phase sensitivity with vacuum state of light as input} and \ref{subsec: Phase sensitivity with squeezed vacuum state of light as input}, we study the phase sensitivities obtained with an HA-enhanced MZI with a coherent state, a vacuum state and a squeezed vacuum state as an input. We investigate the parameter regime in which the phase sensitivities obtained with an HA-enhanced MZI outperform other schemes, beating the SQL and the HL, in Sec.~\ref{sec: Comparison with existing schemes}. Finally, in Sec.~\ref{sec: Phase Sensitivity Enhancement in the Presence of Photon Loss}, we examine the performance of the HA-enhanced MZI sensitivity in the presence of photon losses, which we model through beamsplitters coupled to ancillary modes in the vacuum state. Details of the analytical calculations, numerical simulations, and additional discussions are provided in Appendices \ref{sec: Appendix- Calculation} to \ref{sec: Appendix - Photon loss model}.

\section{Enhancement of MZI sensitivity via Hamiltonian amplification}\label{sec: Enhancement of MZI Sensitivity via Hamiltonian Amplification}

\begin{figure}[ht!]
 \centering
\includegraphics[width = 0.825\columnwidth]{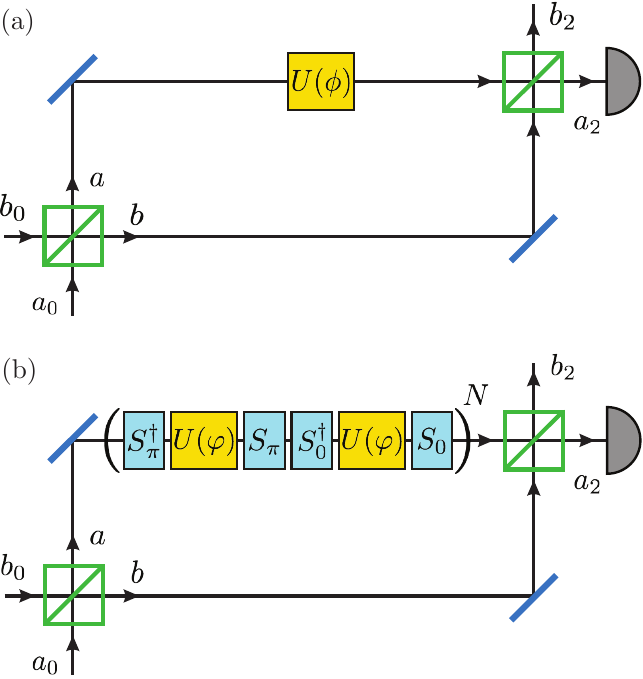}
\caption{Schematic representation of (a) a conventional Mach--Zehnder interferometer and (b) a HA-enhanced Mach--Zehnder interferometer introduced in this work. The phase shift in one arm of the Mach--Zehnder interferometer is described by the phase shifter $U(\phi)$ (yellow). In (b), squeezing transformations of orthogonal phase (blue) are placed equidistantly where $\varphi=\frac{\phi}{2N}$ in one arm of the interferometer. For sufficiently many repetitions $N$ of the squeezing sequence, the phase shift $\phi$ is enhanced by $\cosh(2r)$ where $r$ is the squeezing strength of each squeezer independently of the input states to the interferometer.  In both setups, the phase estimation of $\phi$ is performed via the single-mode intensity detection of the output $a_{2}$ mode.}
\label{fig: MZIschematic}
\end{figure}

A conventional optical MZI comprises a monochromatic source of light, which is split into two beams through a balanced beamsplitter, which, after reflection through plane mirrors, is recombined by a second balanced beamsplitter whose output beams are collected by photon detectors \cite{born2013principles, gerry2023introductory}. As depicted in Fig.~\ref{fig: MZIschematic}(a), the phase shift $\phi$ is described by a phase shifter in one arm of the interferometer given by the unitary transformation $U(\phi) =  \exp(-i \phi a^{\dagger}a)$ where $a$ and $a^{\dagger}$ are the bosonic annihilation and creation operators for the mode $a$. The input modes to the beamsplitter are represented by the annihilation operators $a_0, b_0$, while the output modes of the second beamsplitter are represented by the annihilation operators $a_2, b_2$. Throughout this work, we assume that the phase estimation is performed using intensity detection of the $a_{2}$ mode (\emph{i.e.,} measuring the observable $O = a^{\dagger}_{2}a_{2}$).

Using error propagation theory, the sensitivity or precision in estimating the phase $\phi$ is, for an unbiased estimator, given by \cite{PhysRevA.108.042419,doi:10.1142/S0219749909004839}
\begin{equation}
\label{eq:errorprop}
\Delta \phi = \frac{\Delta O}{ \left| \frac{\partial \langle O \rangle}{\partial \phi} \right|},     
\end{equation}
where $\Delta O = \sqrt{\langle O^{2}\rangle - \langle O \rangle^{2} }$ is the standard deviation of the observable $O$.

\subsection{Hamiltonian amplification}\label{sec: Hamiltonianamplification}

The phase shift described by $U(\phi)$ on one of the arms of a MZI can be amplified by interspersing the phase shifter with local, single-mode squeezing operations, using a protocol called Hamiltonian amplification (HA) \cite{arenz2020amplification}. The phase enhancement through HA can be understood as follows. Consider the single-mode squeezing operator $S_{\theta}(r) = \exp\left[\frac{r}{2}\left(  a^{2}e^{-i\theta}-  a^{\dagger 2}e^{i\theta}\right)\right]$, where $r\geq0$ is the squeezing strength and $\theta \in [0,2\pi)$ is the squeezing angle that describes the quadrature along which the mode is squeezed. Throughout the remainder of this work, we omit the explicit dependence of the squeezing operator on $r$ and write $S_{\theta}(r) \equiv S_{\theta}$. We next consider the sequence of unitaries $S^{\dagger}_{\theta} U(\phi)S_{\theta}$ which, up to a global phase, is equivalent to $\exp(-i\phi H_{(\theta,r)})$, where
\begin{equation}\label{eq:Hamiltonian H_(theta,r)}
  H_{(\theta,r)} =   \cosh(2r) \,  a^{\dagger}a + \frac{\sinh(2r)}{2} \,  (a^{2} e^{-i \theta} + a^{\dagger 2}e^{i \theta})   .
\end{equation}
Thus, as a result of squeezing $S$ and squeezing along the orthogonal quadrature $S^{\dagger}$, which we refer as \emph{anti-squeezing}, the number operator $a^{\dagger}a$ is amplified by the factor $\cosh(2r)$, while additional squeezing-induced terms proportional to $a^{2}$ and $a^{\dagger 2}$ are introduced. However, the additional introduced terms can be suppressed when one alternates between squeezing and anti-squeezing, as described by the Trotterized sequence 
\begin{equation}\label{eq:HAsequence}
\begin{aligned}
U_{N}(\phi) = &\left( S^{\dagger}_{\theta + \pi} U(\varphi)S_{\theta+ \pi} S^{\dagger}_{\theta} U(\varphi)S_{\theta}  \right)^{N}\\
=&\left( S_{\theta} U(\varphi) S^{\dagger}_{\theta}S^{\dagger}_{\theta} U(\varphi)S_{\theta}  \right)^{N},
\end{aligned}
\end{equation}
where $\varphi  =\frac{\phi}{2N}$ and $N$ is the number of Trotter steps. In the limit $N \to \infty$, which we refer to as the Trotter limit, the sequence in Eq.~\eqref{eq:HAsequence} is given by
\begin{align}
U_\infty(\phi) =\lim_{N\to\infty} U_{N}(\phi)= \exp(-i \lambda \phi a^{\dagger}a )
\label{eq:U_inf}
\end{align}
where we introduced the amplification factor
\begin{align}
\lambda=\cosh(2r).
\end{align}
Thus, by rapidly alternating between squeezing along orthogonal quadratures, the phase shift is amplified by $\lambda$, regardless of the input states to the interferometer, as depicted in Fig.~\ref{fig: MZIschematic}(b). Notably, this amplification is achieved without prior knowledge of the phase shift $\phi$. For simplicity, in the rest of this work, we assume $\theta = 0$. 

We show in Appendix \ref{sec: Appendix- Calculation} that in the Trotter limit $(N \rightarrow \infty)$, the minimum phase uncertainty of the HA-enhanced MZI is reduced to

\begin{equation}\label{eq: HA-enahncemnt general}
\Delta\phi_{\rm HA, min}^{(\infty)}
=
\frac{\Delta\varTheta_{\rm min}}{\lambda}.
\end{equation}
Here, $\Delta\phi_{\rm HA,min}^{(\infty)}$ is the minimum phase uncertainty of the
HA-enhanced MZI in the Trotter limit, and
$\Delta\varTheta_{\rm min}$ is the minimum phase uncertainty of a standard
MZI with the same arbitrary input state. Thus, the HA protocol enhances the phase sensitivity of a conventional MZI by a factor of $\lambda$.

In the following sections, we investigate how the dynamical enhancement can be combined with the metrological advantages provided by different input states. In particular, we consider three input states:
\begin{equation}
\begin{aligned}
\ket{\psi_1}
& =
|\alpha\rangle_{a_0}\otimes|0\rangle_{b_0},\\
\ket{\psi_2}
& =
|0\rangle_{a_0}\otimes|0\rangle_{b_0}, \\
\ket{\psi_3}
& =
|\beta\rangle_{a_0}\otimes|\xi\rangle_{b_0},
\end{aligned}
\label{eq:different input states}
\end{equation}
where $\alpha, \beta \in \mathbb{C}$ are coherent state amplitudes and $\xi$ parametrizes a squeezed state (defined in Sec.~\ref{subsec: Phase sensitivity with squeezed vacuum state of light as input}). These input states correspond to coherent-vacuum, vacuum-vacuum, and coherent-squeezed vacuum inputs, respectively.

\subsection{Phase sensitivity with coherent state of light as input} \label{subsec:Phase sensitivity with coherent state of light as input}
\begin{figure}[ht!]
 \centering
\includegraphics[scale = 0.67]{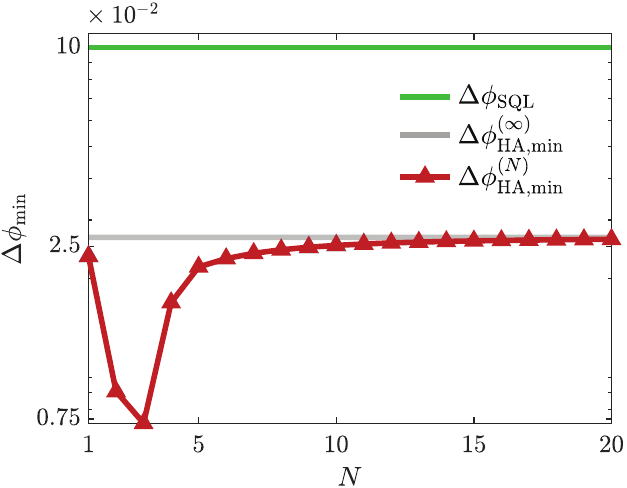}
\caption{Minimum phase uncertainty obtained for the HA-enhanced MZI as a function of the number of Trotter steps $N$ (red) employed in the squeezing sequence given by Eq.~\eqref{eq:HAsequence} for a coherent-vacuum state input defined in Eq.~\eqref{eq:different input states}.
The Trotter limit given in Eq.~\eqref{eq:sensitivity coherent infty} is shown by the grey line. The squeezing strength was set to $r = 1$ and the average number of photons was chosen to be $|\alpha|^{2} = 100$ in the input state. For each Trotter step $N$, the optimal working point $\phi^{*}\in \left(0,\frac{2\pi}{\lambda}\right)$ is chosen such that the phase uncertainty is minimized. For comparison, we also plot the SQL (green) for the same input state.}
\label{fig:phi_min v N coherent}
\end{figure}

The phase sensitivity of a MZI is limited by the SQL when one is restricted to using classical states of light as the input \cite{Woolley_2008}. For instance, when using a coherent state and a vacuum state of light as the input, $\ket{\psi_{1}}$, the phase uncertainty of a MZI is lower bounded by $\Delta \phi_{\rm{SQL}} =  \frac{1}{\sqrt{n}}$, where  $n = |\alpha|^{2}$ is the average number of photons in the coherent state. Using the HA-enhanced MZI, the phase sensitivity can be enhanced in the Trotter limit to
$\Delta \phi^{(\infty)} _{\rm{HA}}(\phi) = \frac{1}{ \lambda\sqrt{n} |\sin( \frac{\lambda \phi}{2} )| }$,
which attains its minimal value
\begin{equation}\label{eq:sensitivity coherent infty}
\Delta \phi^{(\infty)} _{\rm{HA, min}} =\frac{1}{ \lambda\sqrt{n}},
\end{equation}
for an optimal working point $\phi^{*} = (2k+1)\frac{\pi}{\lambda},\,k \in  \mathbb{Z}$,
thereby yielding a minimum uncertainty below the SQL. In Fig.~\ref{fig:phi_min v N coherent} we show the enhancement in the MZI sensitivity beyond the SQL by plotting the minimum uncertainty as a function of the number of Trotter steps implemented in the squeezing sequence given in Eq.~\eqref{eq:HAsequence} with $\ket{\psi_{1}}$ as the input. In the numerical simulations, we set  $r = 1$ and $|\alpha|^{2} = 100$. For each Trotter step $N$ (red triangles), we determine the optimal working point  $\phi^{*}\in \left(0,\frac{2\pi}{\lambda}\right)$ \footnote{The phase uncertainty is generally a function of the phase value $\phi$, and we can define an optical operating point by minimizing the uncertainty over $\phi$: $\Delta\phi(\phi^{*})=\min_{\phi\in(0,2\pi/\lambda)}\Delta\phi(\phi)$. Operationally, this means we bias the interferometer at this optimal working point and detect phase shifts from $\phi^*$}, and the corresponding minimum phase uncertainty is then given by $\Delta\phi_{\rm min}=\Delta\phi(\phi^{*})$. The results suggest that, in the Trotter limit $N\to\infty$ (grey line), an improvement of about $0.6$ orders of magnitude in the minimum phase uncertainty of a MZI over the SQL (green line) can be achieved through the HA-enhanced MZI. We emphasize that, in the Trotter limit, since the dynamics is given by the photon-number conserving unitary, Eq. \ref{eq:U_inf}, the number of photons in the interferometer remains unchanged. The observed improvement in the phase sensitivity is therefore particularly significant as it results from the amplification of the phase shift $\phi$, rather than from an increase in the number of photons. Moreover, for squeezing strengths that yield amplification factors $\lambda > \sqrt{n} $, the sensitivity achieves super-Heisenberg scaling by surpassing the HL $\Delta \phi_{\rm HL}= \frac{1}{n}$, for a total number of photons $n$ in the input state. 

We further observe in Fig.~\ref{fig:phi_min v N coherent} that, surprisingly, the smallest phase uncertainty is obtained for \emph{finitely} many Trotter steps. In particular, the minimum value is achieved for $N = 3$, where we obtain more than an order of magnitude improvement in the phase sensitivity over the SQL. However, we remark that the number of photons in the $N = 3$ case may differ from that considered in the calculation of the SQL. A fair comparison between the two schemes is possible by adopting a common resource measure, which we discuss in Sec.~\ref{sec: Comparison with existing schemes}. For now, we focus on comparing the $N=1$ protocol with the Trotter-limit case. Since the $N=1$ case allows for more tractable analytical calculations, we restrict our attention to this case in the rest of this work and discuss the results for $N \geq 1$ in Appendix \ref{sec: Appendix - N=3}.

\begin{figure}[ht!]
 \centering
\includegraphics[width=\columnwidth]{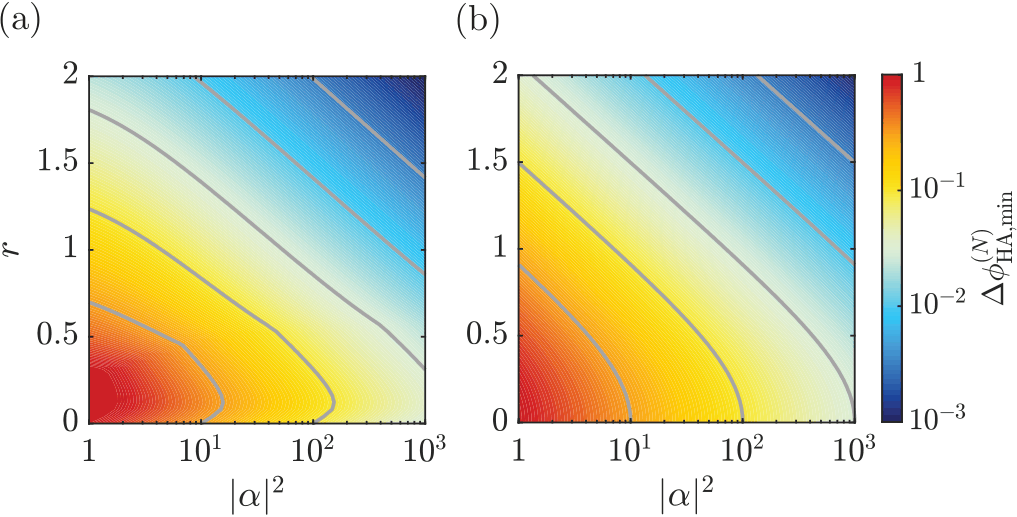}
\caption{Minimum phase uncertainty for the HA-enhanced MZI as a function of the squeezing strength $r$ employed in the amplification sequence given in Eq.~\eqref{eq:HAsequence} and the average number of photons $|\alpha|^{2}$ in the coherent input state $\ket{\psi_{1}}$. The color maps show the minimum phase uncertainty, where in (a) $N=1$ while in (b) $N \to \infty$. }
\label{fig:colormap phi_min coherent}
\end{figure}

In order to further investigate the scaling behavior obtained for a finite number of Trotter steps, we plot the minimum phase uncertainty as a function of both the squeezing strength and the average number of photons in the coherent state $\ket{\psi_{1}}$ in Fig.~\ref{fig:colormap phi_min coherent}. The results are shown in Fig.~\ref{fig:colormap phi_min coherent}(a) for $N=1$ and in Fig.~\ref{fig:colormap phi_min coherent}(b) for the Trotter limit, where the minimum uncertainty is given by the colormap. In both cases, for fixed values of $r$ and $\alpha$, the optimal working point $\phi^{*}\in(0,2\pi/\lambda)$ is selected to minimize the phase uncertainty. The colormaps indicate that for large $|\alpha|^{2}$ and large $r$, the minimum uncertainties obtained with both $N = 1$ and $N \to \infty$ are nearly identical, reaching values $\Delta \phi^{(N)}_{\rm{HA,min}} \lesssim  10^{-3}$. Specifically, for $r \gtrsim 1.5$ and $|\alpha|^{2} \gtrsim 10^{2}$, the minimum uncertainty obtained with $N = 1$ either outperforms or is equivalent to the sensitivities obtained in the Trotter limit. A notable difference between the results in Fig.~\ref{fig:colormap phi_min coherent}(a) and Fig.~\ref{fig:colormap phi_min coherent}(b) can be observed for small values of $|\alpha|^{2}$. In particular, for the range $1\leq |\alpha|^{2} \lesssim 10$ and sufficiently strong $r$, HA with $N =1$ outperforms the Trotter limit case.

\subsection{Phase sensitivity with vacuum state of light as input} \label{subsec:Phase sensitivity with vacuum state of light as input}

\begin{figure}[htbp]
 \centering
\includegraphics[scale = 0.675]{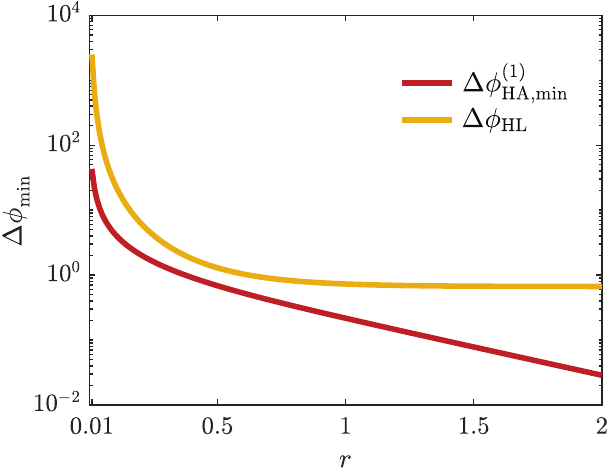}
\caption{Minimum phase uncertainty obtained with the HA-enhanced MZI for $N=1$ with vacuum-vacuum input given in Eq.~\eqref{eq: phasemin vaccum N_1} (red) and the corresponding Heisenberg-limited sensitivity defined in Eq.~\eqref{eq:HL n_out N=1 vacuum} (yellow) as a function of the squeezing strength $r$ employed in the HA sequence. In both cases, the phase uncertainty is calculated at the optimal working point $\phi^{*}$ given in Eq.~\eqref{eq:optimal phase vaccum N1}.}
\label{fig: Appendix- phase sens vacuum vs r vacuum input}
\end{figure}

For the vacuum-vacuum state $\ket{\psi_{2}}$ as the input to a conventional MZI, the phase shift is unobservable since no photons are present in the interferometer. In contrast, it has been shown that replacing the beamsplitters, which preserve the photon number, in a MZI with active elements, such as optical parametric amplifiers or four-wave mixers, enables Heisenberg-limited sensitivity using a vacuum-vacuum input state \cite{PhysRevA.33.4033, PhysRevLett.119.223604, PhysRevA.95.063843}. In these schemes, photons are injected into the interferometer by the active elements. Likewise, the squeezing operations in an HA-enhanced MZI generate photons within the interferometer. However, unlike in the afore mentioned schemes, with $N=1$ and a vacuum-vacuum input state $\ket{\psi_{2}}$, the minimum phase uncertainty 
\begin{equation}\label{eq: phasemin vaccum N_1}
\Delta \phi^{(1)} _{\rm{HA, min}} =   \sqrt{  \frac{ \sqrt {6 \sinh^{2}(4r) + 9} + 3  }{2 \sinh^{2}(4r)} },
\end{equation}
calculated at the optimal working point 
\begin{equation}\label{eq:optimal phase vaccum N1}
\phi^{*}  = \pm 2\tan^{-1} \left(\sqrt{\frac{3}{\sqrt{6 \sinh^{2}(4r) + 9}}}\right) + 2 \pi k, \, k \in \mathbb{Z},
\end{equation}
surpasses the corresponding Heisenberg-limited uncertainty $\Delta \phi_{\text{HL}} = 1/n_{\text{out}}$, given by
\begin{equation}\label{eq:HL n_out N=1 vacuum}
\Delta \phi_{\text{HL}}  =  \frac{\left( \sqrt {6 \sinh^{2}(4r) + 9} + 3  \right)^{2} }{9 \sinh^{2}(4r)},
\end{equation}
hence achieving super-Heisenberg scaling. Here, the output photon number is given by the expectation value of the total photon number operator, i.e., $n_{\text{out}}= \langle a_{2}^{\dagger}a_{2} + b_{2}^{\dagger}b_{2} \rangle$, which, in this case, simplifies to
\begin{equation}\label{eq:n_out N =1 vacuum at phi*}
n_{\text{out}} = \left|\sinh(4r) \sin^{2}\left(\frac{\phi^{*}}{2} \right)\right|^{2}.    
\end{equation}
In Fig.~\ref{fig: Appendix- phase sens vacuum vs r vacuum input} we plot the phase uncertainties given in Eq.~\eqref{eq: phasemin vaccum N_1} and Eq.~\eqref{eq:HL n_out N=1 vacuum} as a function of squeezing strength employed in the HA sequence. These results show that an HA-enhanced MZI achieves super-Heisenberg scaling with a vacuum-vacuum input for any nonzero squeezing strength. In particular, for large squeezing strengths, the uncertainty $\Delta \phi_{\text{HL}}$ (yellow line) saturates to $\approx 0.66$, whereas $\Delta \phi^{(1)} _{\rm{HA, min}}$ (red line) continues to improve with increasing squeezing strength. We analyze the phase sensitivity of the HA-enhanced MZI with a vacuum-vacuum input for different numbers of Trotter steps in Appendix \ref{sec: Appendix - Phase sens with vacuum input}. We further compare the performance of the HA-enhanced MZI with a vacuum-vacuum input state against the SQL and the HL as a function of the total output photon number in Sec.~\ref{sec: Comparison with existing schemes}.

\subsection{Phase sensitivity with squeezed vacuum state of light as input}\label{subsec: Phase sensitivity with squeezed vacuum state of light as input}

In \cite{PhysRevD.23.1693}, C. Caves showed that replacing the vacuum state input of a conventional MZI with a squeezed vacuum state can enhance its phase sensitivity beyond the SQL. In the Caves scheme, the interferometer is initialized in the coherent-squeezed vacuum state $|\psi_{3}\rangle = |\beta\rangle_{a_{0}} \otimes |\xi\rangle_{b_{0}}$, where $\ket{\xi}_{b_{0}}=S_{0}(\xi)\ket{0}_{b_{0}}$. Henceforth, we assume $\xi\in\mathbb{R}^{+}$ and $\beta\in\mathbb{R}$, which is consistent with the choice in \cite{PhysRevA.98.043856} that yields the best sensitivity while simplifying the analysis. The achievable phase sensitivity depends not only on the input state but also on the measurement performed at the interferometer output. To enable a direct comparison with the HA-enhanced MZI, we employ single-detector intensity detection throughout this work, where the intensity is measured at one output port of the interferometer. Although alternative measurement schemes, such as difference intensity detection \cite{PhysRevA.98.043856}, can achieve a lower phase uncertainty, we employ the same single-detector intensity measurement for both the conventional and the HA-enhanced MZI for simplicity and consistency. This isolates the enhancement arising solely from the HA protocol from any improvement due to the measurement strategy. With this measurement scheme, the minimum phase uncertainty of the conventional MZI with coherent-squeezed vacuum input $\ket{\psi_{3}}$ is given by \cite{PhysRevA.98.043856}
\begin{equation}\label{eq:caves_sensitivty_SQL}
\Delta \phi_{\mathrm{Caves,min}}=\frac{\sqrt{\sinh^2 (\xi)+\sqrt{2} \beta \sinh^2 (\xi)+\beta^2 e^{-2 \xi}}}{\left|\beta^2-\sinh ^2 (\xi)\right|}.
\end{equation}
The average input photon number is given by ${n}=|\beta|^{2}+\sinh^{2}(\xi)$, where $|\beta|^{2}$ and $\sinh^{2}(\xi)$ are the average photon numbers in the coherent and squeezed-vacuum inputs, respectively. In the regime $|\beta|^2\gg\sinh^2 (\xi) $, the phase uncertainty given in Eq.~\eqref{eq:caves_sensitivty_SQL} scales as $\mathcal O( \frac{e^{-\xi}}{\sqrt{\beta}})$. Thus, with a squeezed-vacuum state as an input to a MZI, the phase sensitivity of the MZI can surpass the SQL by a factor of $e^{-\xi}$. Notably, when either $\beta = 0$ or $\xi = 0$, the resulting phase sensitivity attains the SQL again. We remark that a single intensity or difference intensity measurement does not fully exploit the phase information available in a MZI with a squeezed input state $\ket{\psi_{3}}$. It was shown later \cite{PhysRevLett.100.073601} that by performing photon number measurements at both output ports and processing the full photon counting statistics using Bayesian phase estimation, the MZI with $\ket{\psi_{3}}$ as input can achieve Heisenberg-limited scaling, $\Delta\phi_{\rm HL} \sim 1/ n$. This scaling is attained in the large photon number regime, $ |\beta|^2, \sinh^2(\xi)\gg1 $, when the coherent and squeezed vacuum inputs contribute approximately equal numbers of photons, $ |\beta|^2 \simeq \sinh^2(\xi) \simeq  n/2$.

For the HA-enhanced MZI, where the induced phase shift is amplified by a factor of $\lambda$, the minimum phase uncertainty in the Trotter limit is
\begin{equation}\label{eq:sensitivity Caves infty}
\Delta \phi^{(\infty)}_{\mathrm{HA,min}}=\frac{\sqrt{\sinh^2 (\xi)+\sqrt{2} \beta \sinh^2 (\xi)+\beta^2 e^{-2 \xi}}}{\lambda \left|\beta^2-\sinh ^2 (\xi)\right|} ,
\end{equation}
which surpasses the SQL even for  $\beta = 0$ or $\xi = 0$. For  $|\beta|^2\gg\sinh^2 (\xi) $, the uncertainty of the HA-enhanced MZI scales as $\mathcal O( \frac{e^{-\xi}}{\lambda |\beta|})$, hence providing an additional enhancement in the phase sensitivity beyond that already achieved by employing a squeezed-vacuum state input. Furthermore, the sensitivity can even surpass the HL when the amplification factor $\lambda$ satisfies 
\begin{equation}
\lambda  > \frac{\beta^2 + \sinh ^2 (\xi)}{\left|\beta^2-\sinh ^2 (\xi)\right| \sqrt{\sinh^2 (\xi)+\sqrt{2} \beta \sinh^2 (\xi)+\beta^2 e^{-2 \xi}}},
\end{equation}
where the HL scaling is here determined by $n = \beta^{2} + \sinh^{2}(\xi)$.

\begin{figure}[ht!]
 \centering
\includegraphics[scale = 0.67]{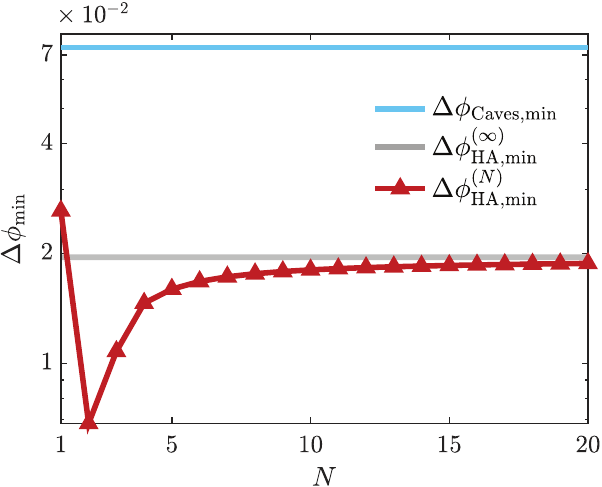}
\caption{Minimum phase uncertainty obtained with the HA-enhanced MZI as a function of the number of Trotter steps $N$ employed in the squeezing sequence given in Eq.~\eqref{eq:HAsequence} (red) for a coherent-squeezed vacuum state defined in Eq.~\eqref{eq:different input states} as the input state. The squeezing strength was set to $r = 1$ in the HA sequence, the average number of photons $|\beta|^{2} = 100$, and the squeezing strength $\xi = 0.5$ in the input state. For comparison, we also plot the Trotter limit shown in Eq.~\eqref{eq:sensitivity Caves infty} (grey) and the minimum uncertainty given by Eq.~\eqref{eq:caves_sensitivty_SQL} (blue) obtained with the assumed input state without employing HA.}
\label{fig:phi_min v N sqz coherent}
\end{figure}

In Fig.~\ref{fig:phi_min v N sqz coherent}, we analyze the phase uncertainty for finitely many Trotter steps. We plot the minimum uncertainty for $|\beta|^{2} = 100$, $\xi = 0.5$ and $r = 1$.  The optimal working point is chosen numerically over the range $\phi^{*}\in \left(0 , \frac{4}{\lambda} \tan^{-1} \left(\sqrt{\frac{\sqrt{2} \beta }{\sinh (2 \xi)}}\right) \right)$. The results in Fig.~\ref{fig:phi_min v N sqz coherent} demonstrate the enhancement in the minimum phase uncertainty obtained with the HA-enhanced MZI $\Delta \phi^{(\infty)}_{\rm{HA, min}} = \Delta \phi_{\rm{Caves, min}}/\lambda$.
Furthermore, the results indicate that, unlike the $\ket{\psi_{1}}$ input, the phase uncertainty obtained with the $N=1$ case does not outperform the results obtained in the Trotter limit for the considered parameters. However, as we increase the number of Trotter steps to $N = 2$, the phase uncertainty $\Delta \phi^{(N)}_{\rm{HA,min}}$ (red triangles) surpasses  $\Delta \phi^{(\infty)}_{\rm{HA,min}}$ (grey line) and reaches its minimum. Thereafter, as we further increase $N$, the minimum phase uncertainty approaches the Trotter limit.

\begin{figure}[ht!]
 \centering
\includegraphics[width=\columnwidth]{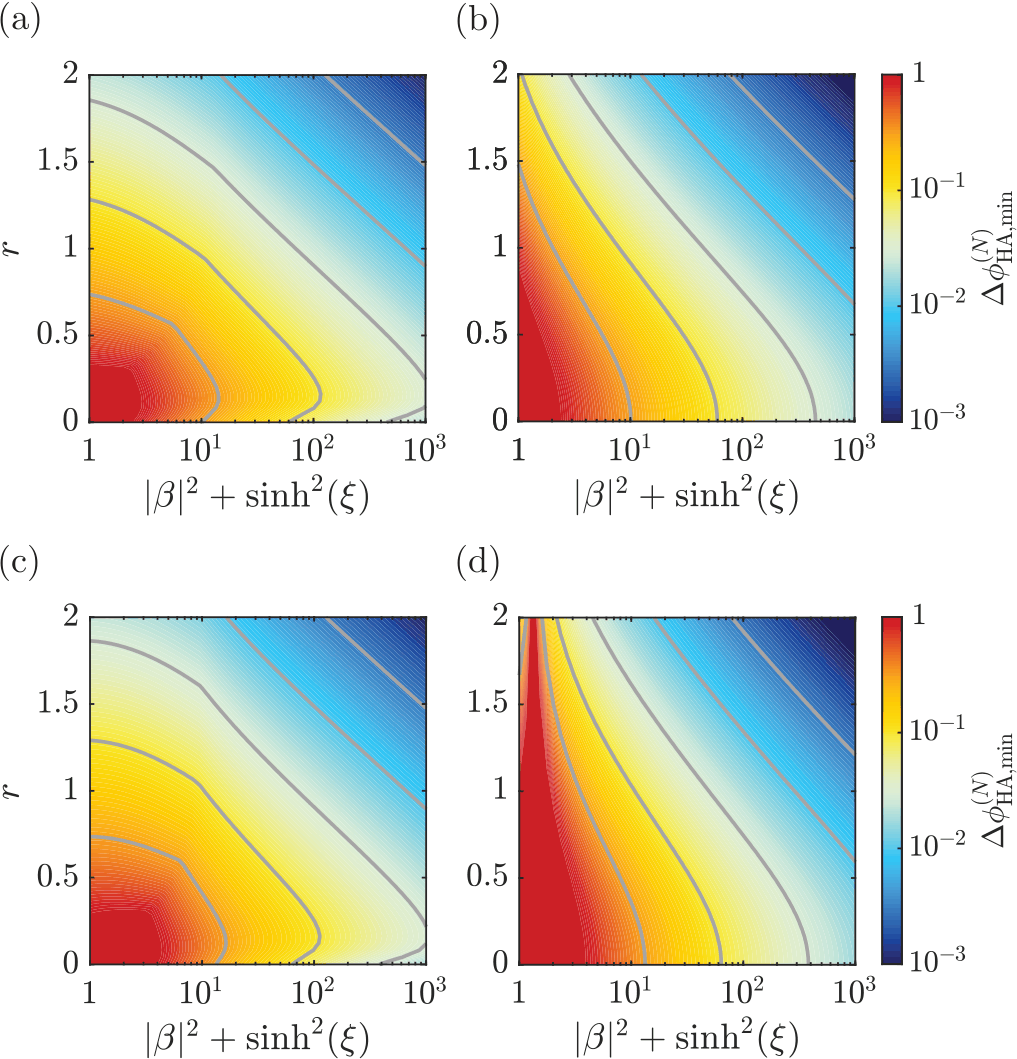}
\caption{Minimum phase uncertainty obtained with the HA-enhanced MZI as a function of the squeezing strength $r$ employed in the amplification sequence defined in Eq.~\eqref{eq:HAsequence} and the average number of photons in the squeezed vacuum - coherent input state given in Eq.~\eqref{eq:different input states}. The color maps show the minimum phase uncertainty for (a), (c) a single Trotter step $N=1$ and (b), (d) the Trotter limit $N\to\infty$. In (a) and (c), we picked $\xi = 0.5$ while for (c) and (d), $\xi = 0.75$ was chosen.}
\label{fig:color maps phi_min v N sqz coherent}
\end{figure}

In Fig.~\ref{fig:color maps phi_min v N sqz coherent}(a) and \ref{fig:color maps phi_min v N sqz coherent}(c) we further examine for the $N = 1$ case the scaling of $\Delta \phi^{(1)}_{\rm{min, HA}}$ as a function of the squeezing strength $r$ and the average number of photons at the input $|\beta|^{2} + \sinh^2(\xi)$. For comparison, we show in Fig.~\ref{fig:color maps phi_min v N sqz coherent}(b) and \ref{fig:color maps phi_min v N sqz coherent}(d) the Trotter limit case. The squeezing strength in the input squeezed states is set to $\xi = 0.5$ in (a) and (b), and $\xi = 0.75$ in (c) and (d). In both cases the optimal working point that yields the minimum phase uncertainty is selected over the range $\phi^{*} \in \left(0 , \frac{4}{\lambda} \tan^{-1} \left(\sqrt{\frac{\sqrt{2} \beta }{\sinh (2 \xi)}}\right) \right)$. The color map plots suggest that for $r \approx 2$ and $|\beta|^{2} \approx 10^{3}$, the minimum phase uncertainties in both cases $N = 1$ and $N \to \infty$ are approximately identical. Furthermore, for a small average number of photons, HA with a single Trotter step outperforms the Trotter limit case at large squeezing strengths. In particular, in the regime $r\gtrsim 1.5$ and $ |\beta|^{2} + \sinh^{2}(\xi) \gtrsim 10^{2}$, the phase sensitivities obtained for a single Trotter step HA-enhanced MZI are comparable to, and can even surpass, the sensitivities obtained in the Trotter limit.

The different phase sensitivities for the different input states are summarized in the Table~\ref{table:comparisontable}

\begin{table*}[htbp]
\centering
\small
\setlength{\tabcolsep}{12pt}
\renewcommand{\arraystretch}{2.3}
\begin{tabular}{c c c c}

\hline\hline

Setup &
Input &
Optimal working point &
Minimum phase uncertainty
\\[1ex]

\hline

MZI
&
$\ket{\alpha}_{a_{0}}\otimes\ket{0}_{b_{0}}$
&
$
\pi
$
&
$
\dfrac{1}{|\alpha|}
$
\\[5ex]

HA-MZI for $N = 1$
&
$\ket{0}_{a_{0}}\otimes\ket{0}_{b_{0}}$
&
$2\tan^{-1} \left(\sqrt{\dfrac{3}{\sqrt{6 \sinh^{2}(4r) + 9}}}\right)
$
&$ \sqrt{  \dfrac{ \sqrt {6 \sinh^{2}(4r) + 9} + 3  }{2 \sinh^{2}(4r)} }$
\\[5ex]

HA-MZI for $N \rightarrow \infty$
&
$\ket{\alpha}_{a_{0}}\otimes\ket{0}_{b_{0}}$
&
$
\dfrac{\pi}{\lambda}
$
&
$
\dfrac{1}
{\lambda |\alpha|}
$
\\[5ex]

MZI
&
$\ket{\beta}_{a_{0}}\otimes\ket{\xi}_{b_{0}}$
&
$
2\tan^{-1}
\left(
\sqrt{
\dfrac{\sqrt2 \beta}
{\sinh(2 \xi)}
}
\right)
$
&
$
\begin{array}{c}
 \mathcal{O}\left( \frac{e^{- \xi}}{|\beta|} \right)^{*}
\qquad
\end{array}
$
\\[-4ex]

&

\\[5ex]

HA-MZI for $N \rightarrow \infty$
&
$\ket{\beta}_{a_{0}}\otimes\ket{\xi}_{b_{0}}$
&
$
\dfrac{2}{\lambda}\tan^{-1}
\left(
\sqrt{
\dfrac{\sqrt2 \beta}{
\sinh(2 \xi)}}
\right)
$
&
$
\begin{array}{c}
\mathcal{O}\left( \frac{e^{- \xi}}{\lambda|\beta|} \right)^{*}
\qquad
\end{array}
$
\\[-4ex]

&

\\[1ex]

\hline\hline

\end{tabular}

\caption{Optimal working point and minimum phase uncertainty of a conventional MZI and an HA-enhanced MZI (HA-MZI) obtained with a single intensity detection scheme for different input states. For coherent-vacuum and coherent-squeezed vacuum inputs given in Eq.~\eqref{eq:different input states}, the average number of photons in the coherent states is given by $|\alpha|^{2}$ and $|\beta|^{2}$, respectively, and $\xi$ corresponds to the squeezing strength used in the generation of the squeezed vacuum state. In all cases, $r$ corresponds to the squeezing strength used in the squeezing sequence defined in Eq.~\eqref{eq:HAsequence} and the amplification factor $\lambda = \cosh(2r)$. $^*$ The asymptotic scaling obtained with a coherent-squeezed vacuum input is obtained for $|\beta|^{2}\gg\sinh^{2}(\xi)$.}

\label{table:comparisontable}

\end{table*}

\section{Further comparison against fundamental uncertainty limits}\label{sec: Comparison with existing schemes}

There are many techniques for beating the SQL and achieving the HL using non-classical states of light \cite{PhysRevD.23.1693,PhysRevLett.85.2733, joo2011quantum, joo2012quantum, bondurant1984squeezed, gerry2010heisenberg, liu2016quantum, campos2003optical,tan2014enhanced,birrittella2012multiphoton}. In this section, we further compare the performance of the HA-enhanced MZI against the SQL and HL. Such a comparison is nontrivial for two reasons. First, the dynamics of the HA-enhanced MZI differ fundamentally from those of a conventional MZI because the HA protocol incorporates squeezing operations within one interferometer arm, which can inject or remove photons during the evolution. Second, the optimal measurement strategy for achieving the best phase sensitivity may differ between the different schemes.

In a conventional MZI, the total photon number is conserved, so the total output photon number is equal to the total input photon number. Consequently, the utilized energy resource is set by the input state and is fixed throughout the interferometric evolution. In contrast, the HA protocol employs dynamical squeezing operations that can modify the photon number, leading to a total energy resource usage and output photon number that depend on the number of Trotter steps $N$, even for a fixed input state. As a result, a direct comparison based solely on the input photon number would not provide a fair assessment of the schemes.

Two resources that are practically constrained in such interferometry schemes are the total intensity of probe light and the total amount of squeezing. Therefore, in the following, we compare schemes based on fixed values of (i) total photon number at the interferometer output, $n_{\text{out}}$, and (ii) total squeezing strength used in the interferometer. 

\begin{figure}[h!]
 \centering
\includegraphics[scale = 0.67]{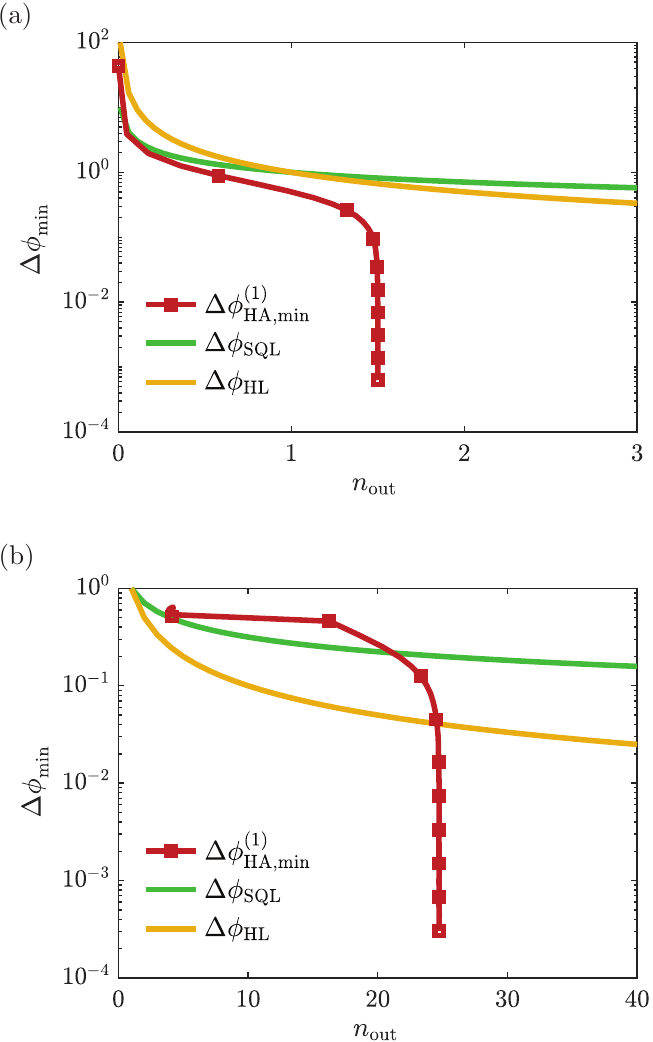}
\caption{Minimum phase uncertainty achieved with the HA-enhanced MZI for $N =1$ (red squares) as a function of the output photon number $n_{\text{out}}$. For comparison, we show the SQL (green) and the HL (yellow). In (a) the vacuum-vacuum state is used as the input, while in (b), the interferometer is initialized in a coherent-vacuum state with $|\alpha|^{2} = 4$. In both cases, the squeezing strength in the HA protocol $(N=1)$ is swept from $r= 10^{-2}$ to $r = 4$, and the corresponding number of photons in the interferometer $n_{\text{out}}$ is calculated. We use the same number of photons to calculate the SQL and the HL.}
\label{fig:phi_min number of photons - comparison}
\end{figure}

For (i) we calculate the total output photons in the HA-enhanced MZI and use it as a common resource measure to determine the corresponding SQL and HL. Here, we restrict ourselves to HA with a single Trotter step $N=1$ and vary the squeezing strength from $r= 10^{-2}$ to $r = 4$. The corresponding output photon number is calculated by $n_{\text{out}}= \langle a_{2}^{\dagger}a_{2} + b_{2}^{\dagger}b_{2} \rangle$, as defined previously. The SQL and HL are then defined as  $\Delta\phi_{\rm SQL}=1/\sqrt{n_{\rm out}}$ and $\Delta\phi_{\rm HL}=1/n_{\rm out}$, respectively \cite{luis2000progress, PhysRevA.55.2598}. The phase sensitivity of the HA-enhanced MZI is subsequently benchmarked against the SQL and HL evaluated at the corresponding output photon number $n_{\rm out}$, \emph{i.e.,} $1/\sqrt{n_{\rm out}}$ and $1/n_{\rm out}$, respectively. This allows us to isolate and quantify the metrological advantage provided by the HA protocol. The results are shown in Fig.~\ref{fig:phi_min number of photons - comparison}, where we plot the phase uncertainty of the HA-enhanced MZI for (a) vacuum-vacuum $\ket{\psi_{2}}$ and (b) coherent-vacuum $\ket{\psi_{1}}$ as the input states. In both Fig.~\ref{fig:phi_min number of photons - comparison}(a) and Fig.~\ref{fig:phi_min number of photons - comparison}(b), each red square corresponds to a distinct squeezing strength and, consequently, a distinct total output photon number. In Fig.~\ref{fig:phi_min number of photons - comparison}(a), the minimum phase uncertainties of the HA-enhanced MZI and the output photon number corresponding to different squeezing strengths are calculated using Eq.~\eqref{eq: phasemin vaccum N_1} and  Eq.~\eqref{eq:n_out N =1 vacuum at phi*}, respectively. 
In Fig.~\ref{fig:phi_min number of photons - comparison}(b), the same two quantities are determined numerically. In both figures, the corresponding SQL and HL are shown by the green and yellow solid lines, respectively. The results in Fig.~\ref{fig:phi_min number of photons - comparison} suggest that the HA-enhanced MZI with a single Trotter step $N=1$ achieves a phase sensitivity that surpasses both the SQL and HL. This enhancement arises because the phase sensitivity of the HA-enhanced MZI increases rapidly, as seen by a sudden drop in Fig.~\ref{fig:phi_min number of photons - comparison} with increasing output photon number, leading to a significantly steeper scaling. The trend towards the saturation of photon numbers originates from the interplay between the optimal phase and the squeezing strength employed in the squeezing sequence. For instance, in the case when $\ket{\psi_{2}}$ is the input state, using the optimal phase shown in Eq.~\eqref{eq:optimal phase vaccum N1}, the total photons at the output saturate to $n_{\text{out}} \approx 1.5$ in the regime $r\gg 1$. Similarly, when $\ket{\psi_{1}}$ is the input state, the total photon number $n_{\text{out}}$ saturates at a finite value and does not increase further with increasing squeezing strength. A detailed analysis of this behavior for $\ket{\psi_{2}}$ and $\ket{\psi_{1}}$ as input states is presented in Appendix \ref{sec: Appendix - Number of output photons v Trotter steps}. We highlight that we selected the range $r \in [10^{-2},4]$ to illustrate the limiting-case behavior. For $r=1.73$ ($\approx 15\,\mathrm{dB}$), close to the highest experimentally demonstrated squeezing, the HA protocol improves upon the SQL and HL by factors of $16.6$ and $13.6$, respectively, for the vacuum-vacuum input state $\ket{\psi_2}$. For the coherent-vacuum input state $\ket{\psi_1}$, the corresponding improvement factors are $8.5$ relative to the SQL and $1.7$ relative to the HL.

\begin{figure}[t!]
 \centering
\includegraphics[scale = 0.67]{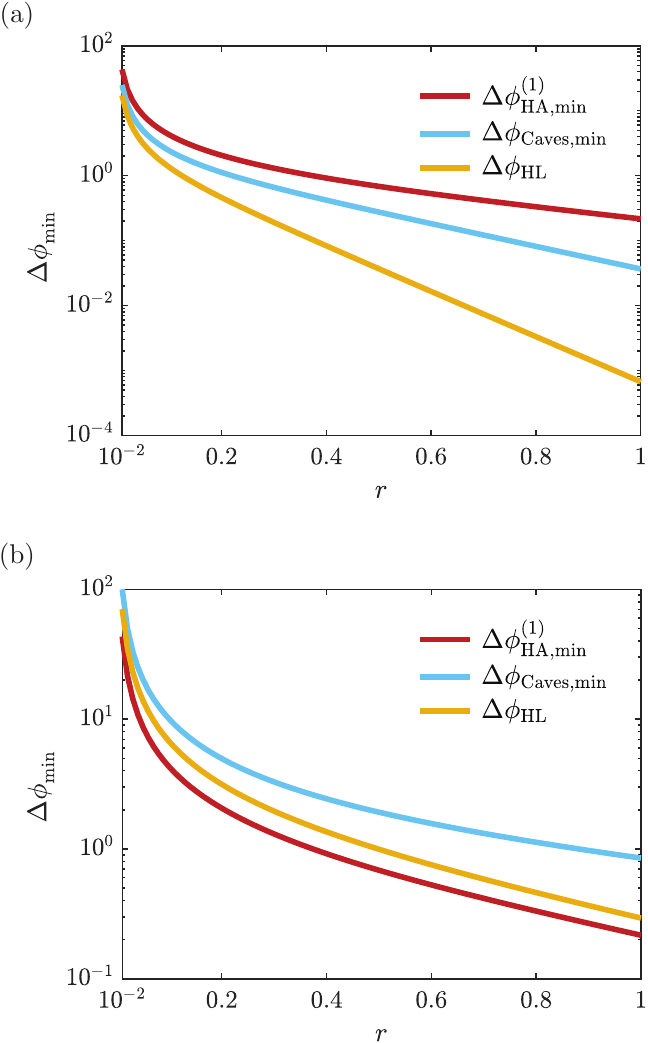}
\caption{Minimum phase uncertainty achieved with a single Trotter step HA-enhanced MZI (red line) as a function of the squeezing strength $r$ for $\ket{\psi_{2}}$ (vacuum-vacuum) input. For comparison, we show the uncertainty of the Cave's scheme (blue line) given by Eq.~\eqref{eq:caves_sensitivty_SQL} and the HL (yellow line) defined following Ref.~\cite{PhysRevLett.100.073601}. The squeezing strength used for the input state $\ket{\psi_{3}}$ for the Caves' scheme was normalized to $\xi=4r$ in (a), while $\xi = r$ is used in (b).}
\label{fig:comparison_based_on_squeezing}
\end{figure}

We go on to consider the case (ii) shown in Fig.~\ref{fig:comparison_based_on_squeezing}, where we adopt the squeezing strength as the common parameter for comparing the different phase-estimation schemes. For the HA-enhanced MZI, we consider a single Trotter step, $N=1$, with the vacuum-vacuum state $\ket{\psi_{2}}$ as the input state and squeezing strength $r$ in the HA sequence. We compare its phase sensitivity with that of the Caves scheme, corresponding to a conventional MZI with the coherent-squeezed vacuum state $\ket{\psi_{3}}$ as the input state, where $\xi$ denotes the squeezing strength of the input squeezed vacuum. In the numerical simulations, we assume $\Delta \phi_{\mathrm{Caves,min}} = 1/|\sinh(\xi)|$, so that the minimum uncertainty achieved with Caves's scheme is equivalent to the SQL.  For the HL, we follow \cite{PhysRevLett.100.073601} and calculate $\Delta \phi_{\mathrm{HL}} = 1/(\abs{\beta}^{2} + \sinh^{2}(\xi))$, where we impose $\sinh^{2}(\xi) = |\beta|^{2}  = n/2$. We consider two scenarios for comparison. In the first scenario, shown in Fig.~\ref{fig:comparison_based_on_squeezing}(a), we normalize the squeezing strengths used in all three schemes by assuming $\xi = 4r$. This follows from the fact that for a single Trotter step, the squeezing sequence given in Eq.~\eqref{eq:HAsequence} contains four squeezing operations, each with squeezing strength $r$, yielding a total squeezing resource of $4r$. In contrast, in Fig.~\ref{fig:comparison_based_on_squeezing}(b), we consider the scenario where $\xi = r$. The results in Fig.~\ref{fig:comparison_based_on_squeezing}(a) indicate that HA fails to yield a metrological advantage when all the schemes are compared under equal squeezing resources. However, the phase sensitivity of the HA-enhanced MZI outperforms both schemes in Fig.~\ref{fig:comparison_based_on_squeezing}(b) in an equal per-squeezer squeezing strengths regime.

\section{Phase Sensitivity Enhancement in the Presence of Photon Losses}\label{sec: Phase Sensitivity Enhancement in the Presence of Photon Loss}

\begin{figure}[h!]
 \centering
\includegraphics[width = 0.95\columnwidth]{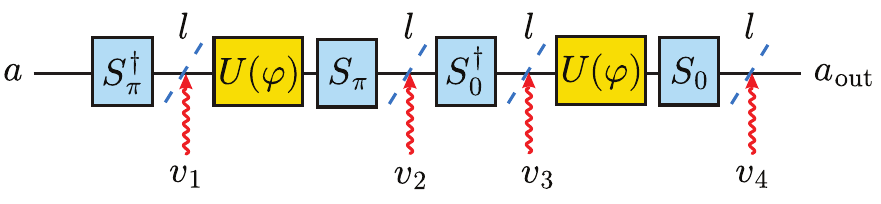}
\caption{Schematic representation of the model used to describe photon losses in an HA-enhanced MZI. Each squeezing operation is followed by a (lossy) beamsplitter (dashed line) containing the loss parameter $l$ that couples the system to auxiliary modes $v_{1},\cdots v_{4}$ prepared in the vacuum state.}
\label{fig:losses}
\end{figure}

We now turn to investigating how photon losses impact the performance of an HA-enhanced MZI. As depicted in Fig.~\ref{fig:losses}, we model losses in the MZI through beamsplitters that couple the system mode $a$ to an auxiliary mode in the vacuum state, where we denote by $l$ the loss coefficient \cite{huang2023optimal,spagnolo2012phase,demkowicz2009quantum}. The loss coefficient is related to the transmissivity $T$ of the lossy beamsplitters via $T = 1-l$. To model squeezing-induced photon loss, we add a lossy beamsplitter after each squeezing transformation in the HA sequence. Thus, for $N$ Trotter steps, there are in total $4N$ lossy beamsplitters. To evaluate the performance of the HA protocol in the presence of photon losses, we consider two scenarios.
First, we consider the case where each lossy beamsplitter in the sequence is assigned a fixed loss parameter $l$ independent of the number of Trotter steps $N$. In this case, the total loss accumulation throughout the interferometer increases very rapidly with increasing $N$.
\begin{figure}[h!]
 \centering
\includegraphics[scale = 0.67]{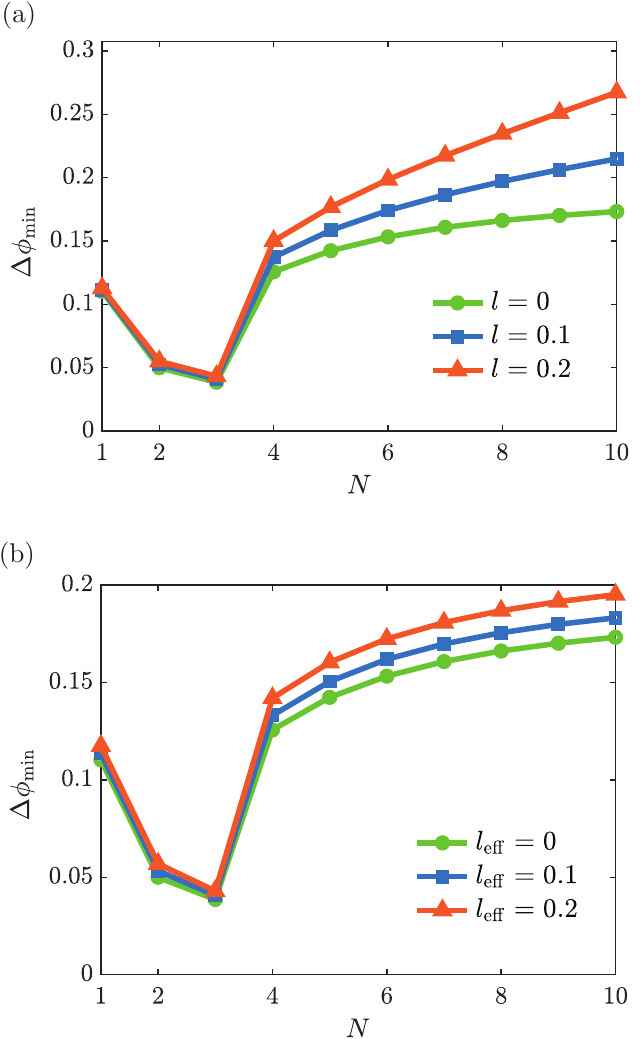}
\caption{Minimum phase uncertainty achieved with the HA-enhanced MZI as a function of the Trotter steps $N$ for coherent-vacuum input $\ket{\psi_{1}}$ with $\alpha = 2$ and the squeezing strength $r=2$ in the sequence given in Eq.~\ref{eq:HAsequence}. In (a), each squeezer has an independent loss parameter $l$, and in (b), the overall loss parameter is fixed at $l_{\rm eff}$.
}
\label{fig:losses_plots}
\end{figure}

In Fig.~\ref{fig:losses_plots}(a) we show the phase uncertainty as a function of Trotter steps $N$ for different values of per beamsplitter loss $l$.  As expected, the accumulation of losses from $4N$ beamsplitters leads to a significant degradation of the phase sensitivity for large Trotter steps $N$. However, the degradation remains relatively modest for small Trotter steps $N$, indicating that the HA protocol retains its robustness against the losses in the low Trotter step regime.  The effect of photon loss for a small number of Trotter steps is further analyzed in Appendix~\ref{sec: Appendix - Photon loss model}, where we also compare the observed loss-robustness to existing schemes. 
In the second scenario, we assume a fixed total loss parameter $l_{\rm eff}$ regardless of the number of Trotter steps.  Consequently, the loss associated with each beamsplitter depends on the number of Trotter steps --- the loss per beamsplitter must be engineered to be smaller for large values of $N$, ensuring that the cumulative effect of all $4N$ lossy beamsplitters reproduces the prescribed total loss $l_{\rm eff}$. More explicitly, suppose the loss per beamsplitter is $l$, then the total transmissivity for $N$ trotter steps becomes $T_{\rm eff} = (1-l)^{4N}$ and the total loss parameter becomes $l_{\rm eff} = (1 - T_{\rm eff }) = 1-(1-l)^{4N}$. This means that each lossy beamsplitter has a loss parameter equal to $l = 1 - (1-l_{\rm eff})^{1/4N}$. 
The resulting phase sensitivities as a function of Trotter steps $N$ for different values of the effective loss parameter $l_{\rm eff}$ are shown in Fig.~\ref{fig:losses_plots}(b). As expected, the phase sensitivity deteriorates with increasing total loss. However, the degradation is still relatively modest for small values of Trotter steps, as in the previous loss model. 

The small deterioration of phase sensitivity with loss for small $N$ shows that the HA protocol retains much of its sensitivity enhancement in the regime where there is the most advantage. As $N$ increases, the accumulated effect of losses becomes more significant, leading to a reduction in performance. The effect of photon loss for a small number of Trotter steps is further examined in Appendix~\ref{sec: Appendix - Photon loss model}, where we also compare the loss robustness of our scheme with that of existing schemes, where the advantages can be considerably more sensitive to losses.

\section{Conclusion}
We introduced a dynamic squeezing protocol that uses squeezing transformation along orthogonal quadratures in one arm of a Mach--Zehnder interferometer to enhance the phase sensitivity of the interferometer, independent of the input states.  We showed that in this way existing quantum sensing strategies that rest on non-classical state inputs can be further improved towards super-Heisenberg scaling. As such, we presented a method that can combine traditional quantum sensing strategies that are based on the preparation of non-classical quantum states with non-linear dynamical evolutions to combine the metrological advantages of both lines of research. We showed that the presented dynamic squeezing protocol is relatively robust under photon loss, thereby making the approach implementable in realistic scenarios. 

We conclude with a discussion of potential experimental realizations of an HA-enhanced MZI. The single-mode dynamic squeezing can be implemented using phase sensitive parametric amplification using a non-linear crystal \cite{10.1119/1.4819195} or in fiber-based MZIs, with nonlinear fiber interactions \cite{Ribeiro:17}. With pulsed pumping, the mode squeezing operations can be interleaved with signal acquisition (\emph{i.e.,} $U(\varphi)$) stages if a timescale separation between these two dynamics can be established. That is, the signal acquisition is likely to be uncontrollable and continuously on, however, if the squeezing operation dominates the dynamics when it is turned on, the Trotterized approximation is valid. While the degree of squeezing per squeezing interaction is likely to limited in this setting due to the pulsed pumping ($r<1$), the number of Trotter steps could be widely varied, and only limited by the interferometer arm length. Identifying low loss non-linear interactions and maintaining phase matching will be the principal challenges with such implementations. Finally, we note that given the similarity between atom interferometry and MZI \cite{PhysRevLett.67.181, RevModPhys.81.1051} it is possible that periodic spin squeezing operations during the time-of-flight portion of the interferometric protocol could enhance the signal sensitivity. We leave it for future work to adapt the Hamiltonian amplification analysis to that setting.

\section*{Acknowledgments}
Sandia National Laboratories is a multimission laboratory managed and operated by National Technology and Engineering Solutions of Sandia LLC, a wholly owned subsidiary of Honeywell International Inc. for the U.S. Department of Energy’s National Nuclear Security Administration under contract DE-NA0003525.

\bibliography{Reference}

\renewcommand{\appendixname}{}

\appendix 
\onecolumngrid

\section*{Appendix}

\noindent
{\hyperref[sec: Appendix- Calculation]{A. \, Enhancement of phase sensitivity in the Trotter limit $N \rightarrow \infty$}
\hfill
\pageref{sec: Appendix- Calculation}

\vspace{0.5em}

\noindent
\hyperref[sec: Appendix - N=3]{B. \, Phase sensitivity of an HA-enhanced MZI with for finite $N$ }
\hfill
\pageref{sec: Appendix - N=3}

\vspace{0.5em}

\noindent
\hyperref[sec: Appendix - Phase sens with vacuum input]{C. \, Phase sensitivity of an HA-enhanced MZI with vacuum input }
\hfill
\pageref{sec: Appendix - Phase sens with vacuum input}

\vspace{0.5em}

\noindent
\hyperref[sec: Appendix - Number of output photons v Trotter steps]{D. \, Number of photons as a function of number of Trotter steps}
\hfill
\pageref{sec: Appendix - Number of output photons v Trotter steps}

\vspace{0.5em}

\noindent
{\hyperref[sec: Appendix - Photon loss model]{E. \, Photon loss model for the HA-enhanced MZI}
\hfill
\pageref{sec: Appendix - Photon loss model}

\vspace{0.5em}

\section{Enhancement of phase sensitivity in the Trotter limit $N \rightarrow \infty$}\label{sec: Appendix- Calculation}
We denote $\Delta\phi_{\rm HA}^{(\infty)}(\phi)$ as the phase uncertainty of the HA-enhanced MZI in the Trotter limit $(N\rightarrow\infty)$. To estimate the unknown phase shift $\phi$, we measure an arbitrary output observable $O$ and employ the standard error propagation formula. In the Trotter limit, the Hamiltonian amplification protocol amplifies the physical phase according to
\begin{equation}
\phi \rightarrow \lambda\phi,
\qquad
\lambda=\cosh(2r),
\end{equation}
so that both the expectation value and the fluctuations of the measurement observable now depend on the amplified phase $\varTheta = \lambda\phi$. Consequently, the phase uncertainty is given by
\begin{equation}
\Delta \phi_{\rm HA}^{(\infty)}(\phi)
=
\frac{\Delta O(\lambda \phi)}
{\left|
\dfrac{\partial}{\partial\phi}
\langle O\rangle(\lambda\phi)
\right|}
=
\frac{\Delta O(\lambda \phi)}
{\lambda
\left|
\dfrac{\partial}{\partial(\lambda\phi)}
\langle O\rangle
\right|},
\label{eq:HA_errorprop}
\end{equation}
which becomes
\begin{equation}
\Delta \phi_{\rm HA}^{(\infty)}(\phi)
=
\frac{\Delta O(\varTheta)}
{\lambda\left|\frac{\partial}{\partial\varTheta}\langle O\rangle\right|}
=
\frac{1}{\lambda}\Delta\varTheta(\varTheta),
\end{equation}
where
\begin{equation}
\Delta\varTheta(\varTheta)
=
\frac{\Delta O(\varTheta)}
{\left|\frac{\partial}{\partial\varTheta}\langle O\rangle\right|}.
\end{equation}
Thus
\begin{equation}
\Delta \phi_{\rm HA}^{(\infty)}(\phi)
=
\frac{1}{\lambda}\Delta\varTheta(\varTheta).
\label{trotter limit phase sensitvity}
\end{equation}
Consequently, estimating the phase $\phi$ in the HA protocol is equivalent to estimating the amplified phase
$\varTheta = \lambda \phi$ in a standard MZI with the same arbitrary input as the HA-enhanced MZI. Notice that $\Delta\phi^{\infty}_{\rm HA}(\phi)$ and $\Delta\varTheta(\varTheta)$ are phase uncertainties evaluated at arbitrary working points; they are not necessarily the minimum uncertainties. To obtain the minimum uncertainty, we must first find the optimal working point. Differentiating the phase uncertainty given by Eq.~\eqref{trotter limit phase sensitvity} with respect to $\phi$ gives
\begin{equation}
\frac{d}{d\phi}\Delta\phi_{\rm HA}^{(\infty)}(\phi)
=
\frac{d}{d\phi}
\left[
\frac{1}{\lambda}\Delta\varTheta(\lambda\phi)
\right]
=0,
\end{equation}
and using
\begin{equation}
\varTheta=\lambda\phi,
\qquad
\frac{d\varTheta}{d\phi}=\lambda,
\end{equation}
yields
\begin{equation}
\frac{d}{d\phi}\Delta\phi_{\rm HA}^{(\infty)}(\phi)
=
\frac{1}{\lambda}
\frac{d\Delta\varTheta}{d\varTheta}
\frac{d\varTheta}{d\phi}
=
\frac{d}{d\varTheta}\Delta\varTheta(\varTheta),
\end{equation}
thus, the optimality condition becomes
\begin{equation}
\frac{d}{d\varTheta}\Delta\varTheta(\varTheta)=0.
\end{equation}
The phase amplified interferometer is optimized when $\varTheta_{\rm opt}=\lambda\phi_{\rm opt}$. Therefore,
\begin{equation}
\phi_{\rm opt}=\frac{\varTheta_{\rm opt}}{\lambda}.
\end{equation}
Finally, the minimum uncertainty becomes
\begin{equation}
\Delta\phi_{\rm HA,min}^{(\infty)}
=
\Delta\phi_{\rm HA}^{(\infty)}(\phi_{\rm opt})
=
\frac{1}{\lambda}
\Delta\varTheta(\lambda\phi_{\rm opt}).
\end{equation}
Using $\varTheta_{\rm opt}=\lambda\phi_{\rm opt}$, the minimum phase uncertainty of the phase-amplified interferometer is related to that of the standard Mach--Zehnder interferometer by
\begin{equation}
\Delta\phi_{\rm HA,min}^{(\infty)}
=
\frac{1}{\lambda}
\Delta\varTheta(\varTheta_{\rm opt})
=
\frac{1}{\lambda}\Delta\varTheta_{\rm min}.
\end{equation}
Hence
\begin{equation}
\Delta\phi_{\rm HA, min}^{(\infty)}
=
\frac{\Delta\varTheta_{\rm min}}{\lambda},
\end{equation}
where $\Delta \varTheta_{\rm min}$ is the minimum phase uncertainty of a standard MZI.

\section{Phase sensitivity of an HA-enhanced MZI with $N\geq1$ }\label{sec: Appendix - N=3}

In this section, we outline the procedure to numerically calculate the phase uncertainty for a finite number of Trotter steps $N$. To efficiently describe the evolution of field operators through the HA protocol, we work in the Heisenberg picture and introduce the operator vector $\mathbf{a}
= \left(a, a^{\dagger} \right)^{T},$ where $a$ denotes the optical mode propagating through the upper arm of the HA-enhanced MZI, in which the Hamiltonian amplification protocol is implemented, as shown in Fig.~\ref{fig: MZIschematic} (b). Since both squeezing and phase-shift operations act linearly on the bosonic operators, their action can be represented by matrix transformations acting on $\mathbf{a}$ \cite{plick2010coherent,DEMKOWICZDOBRZANSKI2015345}. For the single-mode squeezing operator $S_{\theta}(r)
=
\exp\!\left[
\frac{r}{2}
\left(
e^{i\theta}(a^{\dagger})^{2}
-
e^{-i\theta}a^{2}
\right)
\right]$ in the HA sequence 
\begin{equation}
S_{\theta}^{\dagger}(r)\,
\mathbf{a}\,
S_{\theta}(r)
=
\mathbf{S}_{\theta}\,
\mathbf{a},
\end{equation}
where
\begin{equation}
\mathbf{S}_{\theta}
=
\begin{pmatrix}
\mu & e^{i\theta}\nu\\
e^{-i\theta}\nu & \mu
\end{pmatrix},
\label{eq:squeezing matrix}
\end{equation}
with $\mu=\cosh r$ and $\nu=\sinh r$. Similarly, a phase shift $U_{\varphi}=e^{-i\varphi a^{\dagger}a}$  where $\varphi = \phi/2N$, transforms the mode operators as
\begin{equation}
U_{\varphi}^{\dagger}\,
\mathbf{a}\,
U_{\varphi}
=
\mathbf{U}_{\varphi}\,
\mathbf{a},
\end{equation}
where
\begin{equation}
\mathbf{U}_{\varphi}
=
\begin{pmatrix}
e^{-i\varphi} & 0\\
0 & e^{i\varphi}
\end{pmatrix}.
\label{eq:Phase matrix}
\end{equation}
The matrix representation of a single Trotter step is given by
\begin{equation}
\mathbf{U}
=
\mathbf{S}_{0}
\mathbf{U}_{\varphi}
\mathbf{S}_{0}^{\dagger}
\mathbf{S}_{\pi}
\mathbf{U}_{\varphi}
\mathbf{S}_{\pi}^{\dagger}.
\end{equation}
Therefore, the transformation corresponding to \(N\) Trotter steps is given by
\begin{equation}
\mathbf{U}_{N}
=
(\mathbf{U})^{N}
=
\left(
\mathbf{S}_{0}
\mathbf{U}_{\varphi}
\mathbf{S}_{0}^{\dagger}
\mathbf{S}_{\pi}
\mathbf{U}_{\varphi}
\mathbf{S}_{\pi}^{\dagger}
\right)^{N}.
\label{eq:U_N}
\end{equation}

To describe the complete dynamics of the HA-enhanced MZI, we must account for the transformation of interferometer input modes, \(a_{0}\) and \(b_{0}\). We therefore extend the operator vector to include both modes,
\begin{equation}
\mathbf{A_{0}}=\left(\begin{array}{c}
a_{0} \\
a_{0}^{\dagger} \\
b_{0} \\
b_{0}^{\dagger}
\end{array}\right),
\end{equation}
so that every optical element inside the interferometer is represented by a $4 \times 4$ matrix acting on $\mathbf{A_{0}}$. The Hamiltonian amplification sequence only acts on mode $a_{0}$ while the mode $b_{0}$ remains unchanged. The corresponding transformation is therefore
\begin{equation}
{\mathbf{T}}_{N}=\left(\begin{array}{cc}
\mathbf{U}_N & 0 \\
0 & \mathbf{I}_2
\end{array}\right),
\end{equation}
where $\mathbf{U}_{N}$ is given by Eq.~\eqref{eq:U_N} and $\mathbf{I}_{2}$ is a $2\times2$ identity matrix. The two $50:50$ beamsplitters are represented by \cite{plick2010coherent,DEMKOWICZDOBRZANSKI2015345,PhysRevA.33.4033}
\begin{equation}
\mathbf{T}_{\rm BS}=\frac{1}{\sqrt{2}}\left(\begin{array}{cccc}
1 & 0 & i & 0 \\
0 & 1 & 0 & -i \\
i & 0 & 1 & 0 \\
0 & -i & 0 & 1
\end{array}\right).
\end{equation}

Therefore the complete HA-enhanced interferometer transformation maps the input modes $(a_{0}, b_{0})$ into output modes $(a_{2}, b_{2})$ according to  
\begin{equation}
    \mathbf{A}_{2} = \mathbf{T}_{\rm BS} \mathbf{T}_{N}\mathbf{T}_{BS}^{\dagger}\mathbf{A}_{0},
    \label{eq:Total MZI transformation}
\end{equation}
where $\mathbf{T}_{N}$ describes the HA evolution for $N$ Trotter steps.

At the output, we measure the single-detector intensity $O = a_{2}^{\dagger}a_{2}$. The phase uncertainty is estimated using the error propagation formula
\begin{equation}
\label{eq:errorpropsimulations}
\Delta \phi = \frac{\Delta O}{ \left| \frac{\partial \langle O \rangle}{\partial \phi} \right|},     
\end{equation}
where 
\begin{equation}
    \Delta O = \sqrt{\langle O^2 \rangle - \langle O\rangle^{2}}, \quad \langle O\rangle = \bra{\psi}O\ket{\psi}.
    \label{variance raw}
\end{equation}
We consider the input state
\begin{equation}
|\psi_{3}\rangle
=
|\beta\rangle_{a_{0}}
\otimes
|\xi\rangle_{b_{0}},
\label{eq:guassian input}
\end{equation}
where $|\beta\rangle_{a_{0}}$ is a coherent state with complex amplitude $\beta$, and $\ket{\xi}_{b_{0}}=S_{0}(\xi)\ket{0}_{b_{0}}$ is a squeezed vacuum state with squeezing parameter $\xi$. Thus, the average number of photons in the coherent and squeezed-vacuum states is $|\beta|^{2}$ and $\sinh^{2}\xi$, respectively. The coherent-vacuum input $\ket{\psi}_{1} = \ket{\beta}_{a_{0}} \otimes \ket{0}_{b_{0}} $ is recovered by setting $\xi=0$. From Eq.~\eqref{eq:Total MZI transformation}, we see that the measured output has the general form 
\begin{equation}
    a_{2} = Aa_{0} + Ba^{\dagger}_{0} + C b_{0}+ Db^{\dagger}_{0}
    \label{eq:output field},
\end{equation}
where the coefficients $A,B,C,D$ are obtained numerically from the full interferometer transformation given in Eq.~\eqref{eq:Total MZI transformation}. It is convenient to decompose the output field operator into its coherent (mean) amplitude and quantum fluctuation operator \cite{gardiner2004quantum,walls2008quantum,RevModPhys.82.1155},
\begin{equation}
a_{2}
=
\gamma \mathds{1}+\delta a_{2},
\label{eq:field decomposition}
\end{equation}
where
\begin{equation}
\gamma
=
\langle a_{2}\rangle
=
A\beta+B\beta^{*},
\end{equation}
is the coherent amplitude (mean field), and
\begin{equation}
\delta a_{2}
=
a_{2}-\langle a_{2}\rangle,
\label{eq:zero mean}
\end{equation}
describes the quantum fluctuations about the mean field, satisfying
\begin{equation}
\langle\delta a_{2}\rangle=0.
\end{equation}

Physically, this decomposition separates the average coherent output, which carries the phase information, from the quantum fluctuations that ultimately limit the measurement precision. 
More importantly, it greatly simplifies the evaluation of expectation values and variances of output observables by expressing them in terms of the coherent amplitude and the fluctuation moments. Since the fluctuation operator has zero mean, all linear fluctuation terms vanish upon taking expectation values. Furthermore, the Gaussian input state defined in Eq.~\eqref{eq:guassian input}, together with the Gaussian evolution of the interferometer, ensures that the output state remains Gaussian \cite{RevModPhys.84.621}. Consequently, all higher-order fluctuation moments can be expressed in terms of the second-order fluctuation moments through Wick's theorem \cite{walls2008quantum, RevModPhys.84.621}. Therefore, the phase sensitivity can be evaluated entirely from the coherent amplitude and the second-order fluctuation moments. We denote second-order fluctuation moments as
\begin{equation}
n
=
\langle
\delta a_2^\dagger
\delta a_2
\rangle ,
\qquad
m
=
\langle
\delta a_2
\delta a_2
\rangle .
\label{eq:nm_def}
\end{equation}
The remaining second-order moments follow from the bosonic commutation relation \([a_2,a_2^\dagger]=1\) and Hermitian conjugation
\begin{equation}
\langle
\delta a_2
\delta a_2^\dagger
\rangle
=
n+1,
\qquad
\langle
\delta a_2^\dagger
\delta a_2^\dagger
\rangle
=
m^* .
\label{eq:output correlation functions}
\end{equation}
Using Eq.~\eqref{eq:output field}, the fluctuation operator becomes
\begin{equation}
\delta a_2
=
A\delta a_0
+
B\delta a_0^\dagger
+
C\delta b_0
+
D\delta b_0^\dagger .
\end{equation}
From this, the moments become
\begin{equation}
\begin{aligned}
n={}&
|A|^2\langle\delta a_0^\dagger\delta a_0\rangle
+A^*B\langle\delta a_0^\dagger\delta a_0^\dagger\rangle
+A^*C\langle\delta a_0^\dagger\delta b_0\rangle
+A^*D\langle\delta a_0^\dagger\delta b_0^\dagger\rangle \\
&+B^*A\langle\delta a_0\delta a_0\rangle
+|B|^2\langle\delta a_0\delta a_0^\dagger\rangle
+B^*C\langle\delta a_0\delta b_0\rangle
+B^*D\langle\delta a_0\delta b_0^\dagger\rangle \\
&+C^*A\langle\delta b_0^\dagger\delta a_0\rangle
+C^*B\langle\delta b_0^\dagger\delta a_0^\dagger\rangle
+|C|^2\langle\delta b_0^\dagger\delta b_0\rangle
+C^*D\langle\delta b_0^\dagger\delta b_0^\dagger\rangle \\
&+D^*A\langle\delta b_0\delta a_0\rangle
+D^*B\langle\delta b_0\delta a_0^\dagger\rangle
+D^*C\langle\delta b_0\delta b_0\rangle
+|D|^2\langle\delta b_0\delta b_0^\dagger\rangle,
\end{aligned}
\label{eq:n equation}
\end{equation}
and
\begin{equation}
\begin{aligned}
m={}&
A^2\langle\delta a_0\delta a_0\rangle
+AB\langle\delta a_0\delta a_0^\dagger\rangle
+AC\langle\delta a_0\delta b_0\rangle
+AD\langle\delta a_0\delta b_0^\dagger\rangle \\
&+BA\langle\delta a_0^\dagger\delta a_0\rangle
+B^2\langle\delta a_0^\dagger\delta a_0^\dagger\rangle
+BC\langle\delta a_0^\dagger\delta b_0\rangle
+BD\langle\delta a_0^\dagger\delta b_0^\dagger\rangle \\
&+CA\langle\delta b_0\delta a_0\rangle
+CB\langle\delta b_0\delta a_0^\dagger\rangle
+C^2\langle\delta b_0\delta b_0\rangle
+CD\langle\delta b_0\delta b_0^\dagger\rangle \\
&+DA\langle\delta b_0^\dagger\delta a_0\rangle
+DB\langle\delta b_0^\dagger\delta a_0^\dagger\rangle
+DC\langle\delta b_0^\dagger\delta b_0\rangle
+D^2\langle\delta b_0^\dagger\delta b_0^\dagger\rangle.
\end{aligned}
\label{eq:m equation}
\end{equation}
To evaluate \(n\) and \(m\), we use the fact that the two input modes are uncorrelated. Consequently, all second-order cross moments involving both input modes \(a_{0}\) and \(b_{0}\) vanish, and only the second-order moments within each individual mode contribute. We denote these by \(n_{a,b}\) and \(m_{a,b}\). 
For the coherent input state $\ket{\beta}_{a_{0}}$, we write $a_{0}=\beta+\delta a_{0}$, where $\beta=\langle a_{0}\rangle$ and $\langle\delta a_{0}\rangle=0$. The corresponding second-order fluctuation moments are

\begin{equation}
n_a
=
\langle
\delta a_0^\dagger
\delta a_0
\rangle
=
0,
\qquad
m_a
=
\langle
\delta a_0
\delta a_0
\rangle
=
0,
\qquad
\langle
\delta a_0
\delta a_0^\dagger
\rangle
=
1.
\end{equation}

For the squeezed-vacuum input state $\ket{\xi}_{b_{0}}$, writing $b_{0}=\delta b_{0}$ with $\langle\delta b_{0}\rangle=0$, we obtain the second-order fluctuation moments
\begin{equation}
\langle
\delta b_0^\dagger
\delta b_0
\rangle
=
\sinh^2\xi,
\qquad
\langle
\delta b_0
\delta b_0
\rangle
=
-\sinh\xi\cosh\xi,
\qquad
\langle
\delta b_0
\delta b_0^\dagger
\rangle
=
\cosh^2\xi.
\end{equation}
Substituting the input second-order fluctuation moments into Eq.~\eqref{eq:n equation} and Eq.~\eqref{eq:m equation}, we obtain
\begin{equation}
\begin{aligned}
n ={}&
|B|^2 + |C|^2 n_{b} + |D|^2(n_{b}+1)
+ 2\mathrm{Re}(C^{*}Dm_{b}),
\end{aligned}
\label{eq:n final}
\end{equation}
and
\begin{equation}
\begin{aligned}
m ={}&
AB + C^2m_{b} + D^2m_{b}^{*}
+ CD(2n_{b}+1).
\end{aligned}
\label{eq:m final}
\end{equation}
Using the decomposition shown in Eq.~\eqref{eq:field decomposition} the measured observable becomes
\begin{equation}
O=a_{2}^{\dagger}a_{2}
=|\gamma|^2+\gamma^*\delta a_{2}+ \gamma \delta a_{2}^{\dagger}+\delta a_{2}^{\dagger}\delta a_{2},
\label{eq:observable}
\end{equation}
and taking the expectation value and using $\langle \delta a_{2} \rangle = 0$, we find
\begin{equation}
\begin{aligned}
    \langle O \rangle & = |\gamma|^2 + \langle \delta a^{\dagger}_{2} \delta a_{2} \rangle \\
    & =|\gamma|^2 + n.
\end{aligned}
\label{eq:O average}
\end{equation}
Now, from the Eq.~\eqref{eq:observable} the square of the measured observable becomes
\begin{equation}
\begin{aligned}
O^2= & |\gamma|^4+|\gamma|^2 \gamma^* \delta a_{2}+|\gamma|^2 \gamma \delta a^{\dagger}_{2}+|\gamma|^2 \delta a^{\dagger}_{2} \delta a_{2} +|\gamma|^2 \gamma^* \delta a_{2}+\left(\gamma^*\right)^2 \delta a_{2} \delta a_{2}+|\gamma|^2 \delta a_{2} \delta a^{\dagger}_{2} \\
& +\gamma^* \delta a_{2} \delta a^{\dagger}_{2} \delta a_{2}
+|\gamma|^2 \gamma \delta a^{\dagger}_{2}+|\gamma|^2 \delta a^{\dagger}_{2} \delta a_{2} +\gamma^2 \delta a^{\dagger}_{2} \delta a^{\dagger}_{2}
+\gamma \delta a^{\dagger}_{2} \delta a^{\dagger}_{2} \delta a_{2} \\
& +|\gamma|^2 \delta a^{\dagger}_{2} \delta a_{2}
+\gamma^* \delta a^{\dagger}_{2} \delta a_{2} \delta a_{2} +\gamma \delta a^{\dagger}_{2} \delta a_{2} \delta a^{\dagger}_{2}
+\delta a^{\dagger}_{2} \delta a_{2} \delta a^{\dagger}_{2} \delta a_{2}.
\label{eq: operator square}
\end{aligned}
\end{equation}
Since $\delta a_{2}$ is a zero-mean Gaussian field from Eq.~\eqref{eq:zero mean}, Wick's theorem \cite{walls2008quantum, RevModPhys.84.621} can be applied to reduce the fourth-order moments to a product of second-order moments when we take the average. In particular, 
\begin{equation}
\begin{aligned}
     \langle \delta a_{2}^{\dagger} \delta a_{2}^{\dagger} \delta a_{2} \delta a_{2} \rangle
     &= \langle \delta a_{2}^{\dagger} \delta a_{2}^{\dagger} \rangle
        \langle \delta a_{2} \delta a_{2} \rangle
      + \langle \delta a_{2}^{\dagger} \delta a_{2} \rangle
        \langle \delta a_{2}^{\dagger} \delta a_{2} \rangle \\
    &\quad
      + \langle \delta a_{2}^{\dagger} \delta a_{2} \rangle
        \langle \delta a_{2}^{\dagger} \delta a_{2} \rangle \\
     & = |m|^2 + 2|n|^2 .
    \label{wicks_theorem}
\end{aligned}
\end{equation}
Using Eq.~\eqref{eq:nm_def} and Eq.~\eqref{eq:output correlation functions} the expectation value of the square of observable becomes
\begin{equation}
\begin{aligned}
\langle O^2\rangle & =
|\gamma|^4
+|\gamma|^2(4n+1)
+2{\rm Re}\!\left[(\gamma^*)^2m\right] \\
& +|m|^2
+2n^2
+n .
\end{aligned}
\end{equation}
Therefore, the variance given in Eq.~\eqref{variance raw} becomes
\begin{equation}
\label{eq:variance_final}
\Delta O
=
\sqrt{
|\gamma|^{2}(1+2n)
+2\,\mathrm{Re}\!\left[(\gamma^{*})^{2}m\right]
+n(n+1)
+|m|^{2}
},
\end{equation}
and by substituting Eq.~\eqref{eq:O average} and Eq.~\eqref{eq:variance_final} into Eq.~\eqref{eq:errorpropsimulations}, the phase uncertainty becomes
\begin{equation}
\Delta\phi
=
\frac{
\sqrt{
|\gamma|^2(1+2n)
+
2{\rm Re}\!\left[(\gamma^*)^2m\right]
+
n(n+1)
+
|m|^2
}
}{
\left|
\dfrac{\partial}{\partial\phi}
\left(
|\gamma|^2+n
\right)
\right|
}.
\label{eq:phase_sensitivity_general}
\end{equation}
Here, \(n\) and \(m\) are given by Eq.~\eqref{eq:n final} and Eq.~\eqref{eq:m final}, respectively. Since the coefficients \(A\), \(B\), \(C\), and \(D\) generally do not admit closed-form expressions for arbitrary Trotter steps \(N\), they are obtained numerically from
Eq.~\eqref{eq:Total MZI transformation}. The minimum phase uncertainty is then calculated from Eq.~\eqref{eq:phase_sensitivity_general}, by choosing the optimal working point $\phi^{*}\in(0,2\pi/\lambda)$, where $\lambda=\cosh(2r)$, such that $\Delta\phi(\phi^{*})=\min_{\phi\in(0,2\pi/\lambda)}\Delta\phi(\phi)$. The corresponding minimum phase uncertainty is then given by $\Delta\phi_{\rm min}=\Delta\phi(\phi^{*})$. This gives the minimum phase uncertainty for an HA-enhanced MZI with coherent-squeezed vacuum input $\ket{\psi_{3}}$. 
For coherent-vacuum input $\ket{\psi_{1}} = \ket{\beta}_{a_{0}}\otimes\ket{0}_{b_{0}}$, the squeezed vacuum reduces to the ordinary vacuum state (\(\xi=0\)), implying 
\begin{equation}
n_b=0,
\qquad
m_b=0.
\end{equation}
Substituting these into Eq.~\eqref{eq:n final} and Eq.~\eqref{eq:m final} yields
\begin{equation}
n
=
|B|^2+|D|^2,
\label{coherent n}
\end{equation}
and
\begin{equation}
m
=
AB+CD.
\label{coherent m}
\end{equation}
The phase uncertainty for the coherent-vacuum input state $\ket{\psi_{1}}$ is then obtained from Eq.~\eqref{eq:phase_sensitivity_general} using Eq.~\eqref{coherent n} and Eq.~\eqref{coherent m}.
\section{Phase sensitivity of an HA-enhanced MZI with vacuum input}\label{sec: Appendix - Phase sens with vacuum input}
In this section, we analyze the minimum phase uncertainty obtained through an HA-enhanced MZI with vacuum-vacuum input. In Fig.~\ref{fig: Appendix- phase sens vacuum vs r finite N} we plot the minimum phase uncertainties for different number of Trotter steps employed in the squeezing sequence as a function of the squeezing strength $r$. The results indicate that as the number of Trotter steps is increased from $N = 1$ (black) to $N = 2$ (blue) and then to $N =3$ (red), the phase uncertainty achieves its minimum value. 
As the number of Trotter steps is increased to $N = 4 $ (green), $N = 5 $ (yellow), $N = 6 $ (purple), and $N = 10$ (violet), the phase uncertainty increases. This behavior is expected, as the phase uncertainty of the HA-enhanced MZI with vacuum–vacuum input is infinitely large in the Trotter limit $N \to \infty$.
\begin{figure}[H]
 \centering
\includegraphics[scale = 0.675]{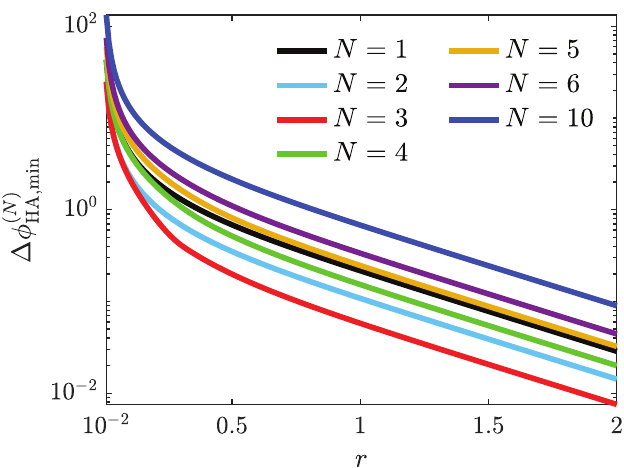}
\caption{Minimum phase uncertainty achieved with the HA-enhanced MZI with vacuum-vacuum input as a function of the squeezing strength $r$ for different number of Trotter steps $N = 1$ (black), $N = 2$ (blue), $N = 3$ (red), $N = 4$ (green), $N = 5$ (yellow), $N = 6$ (purple) and $N = 10$ (violet). The minimum uncertainty for $N = 1$ is calculated using Eq.~\eqref{eq: phasemin vaccum N_1}, whereas phase uncertainty for the rest of the cases is calculated numerically. In the numerical calculations, the optimal angle $\phi^{*}$ is selected over the range $\left( 0, \frac{2\pi}{\lambda}\right)$, where $\lambda =\cosh(2r)$.}
\label{fig: Appendix- phase sens vacuum vs r finite N}
\end{figure}

\section{Number of photons as a function of the number of Trotter steps}\label{sec: Appendix - Number of output photons v Trotter steps}

 In this section, we investigate how the total output photon number in the HA-enhanced MZI varies with the number of Trotter steps across a range of squeezing strengths. In Fig.~\ref{fig: Appendix -Total_photon_vs_N}(a) we plot total output photons $n_{\text{out}}$ for vacuum-vacuum input as described in Eq.~\eqref{eq:n_out N =1 vacuum at phi*}, calculated at the optimal phase defined in Eq.~\eqref{eq:optimal phase vaccum N1}, as a function of the number of Trotter steps $N$. As we expected, on increasing $N$, the total number of photons reduces to $n_{\text{out}} = 0$. In contrast, photon numbers at $N=1$ yields $n_{\text{out}} > 0$ which saturates to $n_{\text{out}} = 1.5$ as we increase squeezing strength from $r= 1$ (blue circle), $r = 2$ (red square), $r= 3$ (yellow diamond) to $r= 4$ (green triangle). We observe a similar behavior in Fig.~\ref{fig: Appendix -Total_photon_vs_N}(b) when coherent-vacuum state $\ket{\psi_{1}}$ is used as an input to the HA-enhanced MZI, where we use $|\alpha|^{2} = 4$. The results suggest that the output photon number saturates to $n_{\text{out}} \approx 25$ when squeezing strength is increased from $r = 1$ (blue circle) to $r = 4$ (green triangle) with $N = 1$, and then reduces to $n_{\text{out}} = 4$ as we increase $N$. We remark that, in contrast to the vacuum-vacuum input case, the $n_{\text{out}}$ does not decrease monotonically with increasing $N$ for a coherent-vacuum input.

\begin{figure}[H]
 \centering
\includegraphics[scale = 0.675]{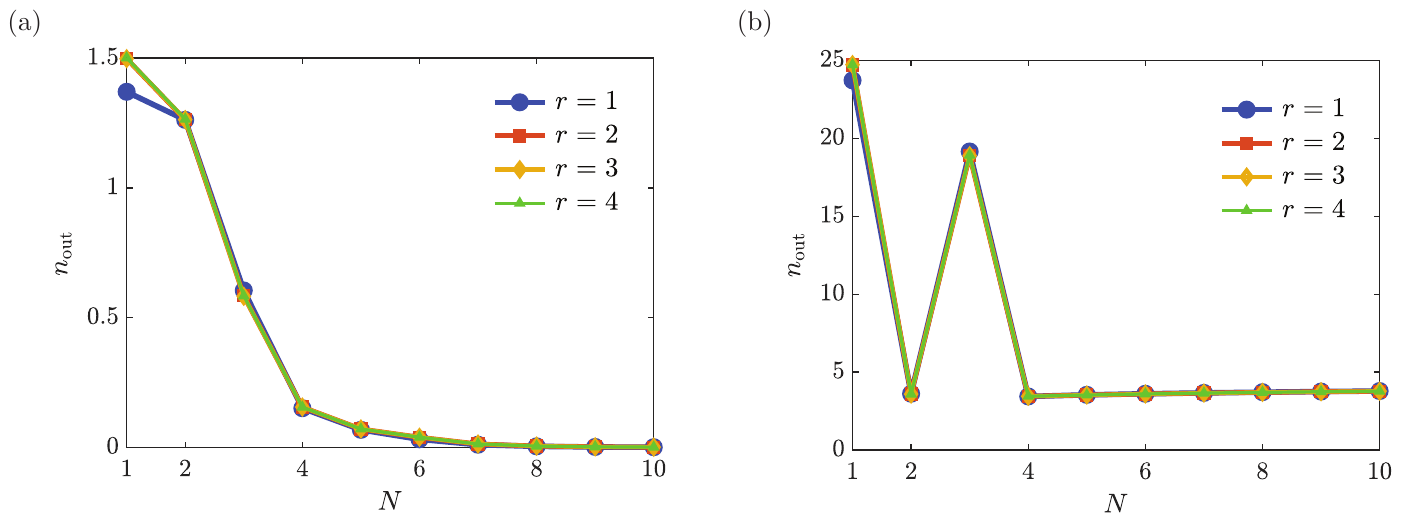}
\caption{Total number of photons in the HA-enhanced MZI as a function of the number of Trotter steps implemented in the squeezing sequence shown in Eq.~\eqref{eq:HAsequence} with (a) vacuum and (b) coherent-vacuum state, as an input to the interferometer. In both cases, the number of photons is plotted for the squeezing strengths $r= 1$ (blue circles), $r = 2$ (red squares), $r =3$ (yellow diamonds), and $r =4$ (green triangles). In the numerical simulations for (b), we set $|\alpha|^{2} = 4$.}
\label{fig: Appendix -Total_photon_vs_N}
\end{figure}

\section{Photon loss model for the HA-enhanced MZI}\label{sec: Appendix - Photon loss model}

The loss model is shown in Fig.~\ref{fig:losses_sequence} where losses are added after every squeezing sequence. In a conventional interferometer, photon losses commute with linear phase accumulation \cite{DEMKOWICZDOBRZANSKI2015345, demkowicz2009quantum}. Consequently, the beamsplitter model is equivalent regardless of where the loss is placed within the phase sensing stage. This is not true for squeezing operations, which do not commute with photon loss. Therefore, in the Hamiltonian amplification protocol, losses must be incorporated after each squeezing operation to accurately account for the losses resulting in degradation of phase sensitivity.
\begin{figure}[ht!]
 \centering
\includegraphics[width = 0.5\textwidth]{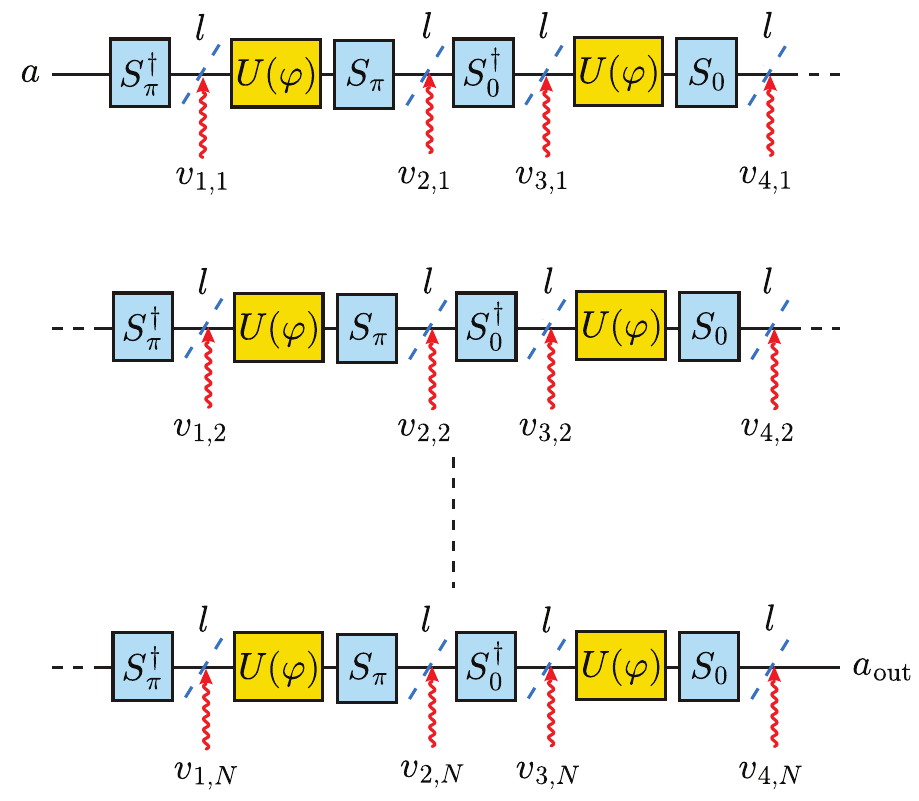}
\caption{Model of losses in the HA-enhanced MZI. Each Trotter step consists of the operations $S_{\pi}^{\dagger}$, $U(\varphi)$, $S_{\pi}$, $S_{0}^{\dagger}$, $U(\varphi)$, and $S_{0}$. Losses are incorporated after every squeezing operation using fictitious beamsplitters (dashed line) with loss parameter $l$, which couple the signal mode to independent environmental vacuum modes $v_{i,j}$. Here, $i=1,\ldots,4$ labels the lossy beamsplitters within a Trotter step, while $j=1,\ldots,N$ labels the Trotter step. After $N$ Trotter steps, the output mode is denoted by $a_{\rm out}$.
}
\label{fig:losses_sequence}
\end{figure}
Since each sequence contains four loss processes, a total of $4N$ lossy beamsplitters are introduced for $N$ Trotter steps. Each loss element couples the interferometer mode to a fresh vacuum mode. To describe the complete evolution, we enlarge the operator space to include both physical and environmental degrees of freedom:
\begin{equation}
\mathbf{A}_{2}=\mathbf{T}_{\rm BS}\mathbf{T}_N^{(l)} \mathbf{T}_{\rm BS}^{\dagger}\mathbf{A}_{0},
\label{loss_model_equation}
\end{equation}
where
\begin{equation}
\mathbf{A}=
\left(
a,\,
a^{\dagger},\,
b,\,
b^{\dagger},\,
v_{1,1},\,
v_{2,1}^{\dagger},\,
\ldots,\,
v_{3,N},\,
v_{4,N}^{\dagger}
\right)^{T}.
\end{equation}
and
\begin{equation}
\mathbf{T}_N^{(l)}
=
\prod_{j=1}^{N}
\left[
\mathbf{L}_{4,j}
\mathbf{S}_{0}^{\dagger}
\mathbf{U}_{\varphi}
\mathbf{L}_{3,j}
\mathbf{S}_{0}
\mathbf{L}_{2,j}
\mathbf{S}_{\pi}^{\dagger}
\mathbf{U}_{\varphi}
\mathbf{L}_{1,j}
\mathbf{S}_{\pi}
\right],
\end{equation}
where $\mathbf{T}_{\rm BS}$ is the beamsplitter transformation and $\mathbf{T}_N^{(l)}$ denotes the total evolution matrix containing squeezing, phase accumulation, and all lossy beamsplitters. Here, $\mathbf{L}_{i,k}$ notation defines the loss after $i\rm th$ squeezing operation in the $k$th Trotter step. Each individual lossy beamsplitter $\mathbf{L}_{i,k}$  is described by a beamsplitter transformation that acts only on the interferometer mode $a$ and its corresponding vacuum mode $v_{i,k}$ described by               
\begin{equation}
\left(\begin{array}{c}
a \\
a^{\dagger} \\
v_{i, j} \\
v_{i, j}^{\dagger}
\end{array}\right) \rightarrow \mathbf{L}_{i, j}\left(\begin{array}{c}
a \\
a^{\dagger} \\
v_{i, j} \\
v_{i, j}^{\dagger}
\end{array}\right),
\end{equation}
with
\begin{equation}
\mathbf{L}_{i, j}=\left(\begin{array}{cccc}
\sqrt{1-l} & 0 & \sqrt{l} & 0 \\
0 & \sqrt{1-l} & 0 & \sqrt{l} \\
-\sqrt{l} & 0 & \sqrt{1-l} & 0 \\
0 & -\sqrt{l} & 0 & \sqrt{1-l}
\end{array}\right),
\end{equation}
where $1-l$ is the transmissivity and $l$ is the loss parameter. From Eq.~\eqref{loss_model_equation}, the general form of the measured output becomes
\begin{equation}
\begin{aligned}
    a_{\rm 2} & = A a_{0} + B a_{0}^{\dagger} + Cb_{0} + Db_{0}^{\dagger} + \sum_{j=1}^{N} \sum_{i=1}^{4} (U_{i,j} v_{i,j} + V_{i,j} v_{i,j}^{\dagger}),
   \label{eq:output field operator}
\end{aligned}
\end{equation}
where \(A\), \(B\), \(C\), \(D\), \(U_{i,j}\), and \(V_{i,j}\) are the complex coefficients determined by the total linear transformation matrix given in Eq.~\eqref{loss_model_equation}. Here, \(a_{0}\) and \(b_{0}\) denote the input mode operators of the HA-enhanced MZI, while \(v_{i,j}\) are the vacuum mode operators introduced by the beamsplitters that model photon losses. The index  $j=1,\ldots,N$ labels the Trotter step, whereas the index  $i=1,\ldots,4$ labels the four beamsplitters within each Trotter step. Here we consider the coherent-vacuum input state $\ket{\psi_1}$. Since the auxiliary modes $v_{i,j}$ correspond to vacuum environmental modes, the total input state is
\begin{equation}
\ket{\psi}
=
\ket{\psi_{1}}
\otimes
\prod_{i,j}\ket{0}_{v_{i,j}},
\label{eq:input modes with vacuum modes}
\end{equation}
where $\ket{\psi_{1}} = \ket{\alpha}_{a_{0}} \otimes\ket{0}_{b_{0}}$. Following the same procedure as in Appendix~\ref{sec: Appendix - N=3}, and using the fact that
\(\langle v_{i,j} \rangle = 0\), we obtain the same expression for the phase sensitivity as given in Eq.~\eqref{eq:phase_sensitivity_general}.
\begin{equation}
\Delta\phi
=
\frac{
\sqrt{
|\gamma|^2(1+2n)
+
2{\rm Re}\!\left[(\gamma^*)^2m\right]
+
n(n+1)
+
|m|^2
}
}{
\left|
\dfrac{\partial}{\partial\phi}
\left(
|\gamma|^2+n
\right)
\right|
}.
\label{eq:phase_sensitivity_general_losses}
\end{equation}
However, in the presence of photon losses, an additional $4N$ vacuum modes are introduced. Consequently, the quantities $n$ and $m$ defined in Eq.~\eqref{eq:nm_def} become
\begin{equation}
    n = |B|^2 + |D|^2 + \sum_{j=1}^{N} \sum_{i=1}^{4} |V_{i,j}|^2 
    \label{eq:n},
\end{equation}
and 
\begin{equation}
    m = AB + CD + \sum_{j=1}^{N} \sum_{i = 1}^{4} U_{i,j} V_{i,j}
    \label{eq:s}.
\end{equation}
Substituting these expressions into Eq.~\eqref{eq:phase_sensitivity_general_losses} yields the phase sensitivity of the HA-enhanced MZI in the presence of losses. 
Having established the general formalism, we now focus on the low Trotter step regime. We saw from Fig.~\ref{fig:losses_plots} that degradation of the phase sensitivity is modest for small Trotter steps. To analyze the HA protocol for the low Trotter step regime $N$ we analyze the sensitivities for  $N=1$ and $N=3$ as a function of the effective loss parameter $l_{\rm eff}$ defined in Sec.~\ref{sec: Phase Sensitivity Enhancement in the Presence of Photon Loss}. As shown in Fig.~\ref{fig:HA-enhanced MZI losses with leff}, the phase sensitivities for both $N=1$ and $N=3$ remain very robust against losses over a broad range of $l_{\rm eff}$. Only in the limit of complete photon loss ($l_{\rm eff} = 1$), the phase sensitivity diverges, as expected since no information about the phase can be extracted in this case.
\begin{figure}[H]
 \centering
\includegraphics[scale = 0.675]{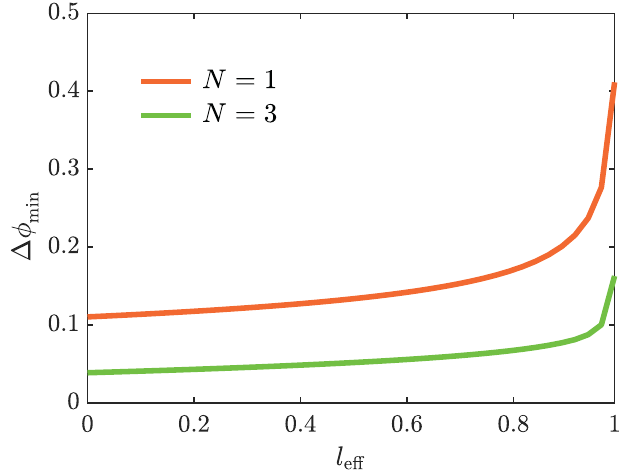}
\caption{ Minimum phase uncertainty $\Delta\phi_{\rm min}$ as a function of the effective loss parameter $l_{\mathrm{eff}}$ for the HA-enhanced MZI with $N=1$ and $N=3$ Trotter steps. The input state is the coherent-vacuum state $\ket{\psi_{1}}$, with squeezing parameter $r=1$ and coherent amplitude $\alpha=2$.}
\label{fig:HA-enhanced MZI losses with leff}
\end{figure}
To compare the loss tolerance of the HA protocol with that of conventional interferometric schemes, we consider a standard Mach--Zehnder interferometer with coherent-vacuum input and a NOON state-based phase estimation protocol \cite{dowling2008quantum}. For a consistent comparison, losses are assumed to occur only in one arm of each interferometer and are modeled by a fictitious beamsplitter with loss parameter $l$. The corresponding phase sensitivities as functions of the loss parameter are shown in Fig.~\ref{fig:Standard MZI with coherent losses}. For a conventional MZI with coherent-vacuum input $\ket{\psi} = \ket{\alpha}_{a_{0}} \otimes \ket{0}_{b_{0}}$, the minimum phase uncertainty in the presence of losses is given by $\Delta \phi_{\rm min} = \frac{1+\sqrt{1-l}}{2\sqrt{1-l}|\alpha|}$ \cite{huang2023optimal}. The dependence of this phase uncertainty on the loss parameter $l$ is shown in Fig.~\ref{fig:Standard MZI with coherent losses}. For the NOON state $\ket{\psi_{\rm NOON}}=\frac{1}{\sqrt{2}}(\ket{0}_{a} \otimes \ket{n}_{b}+\ket{n}_{a} \otimes\ket{0}_{b})$, the phase uncertainty in the presence of losses is given by $\Delta \phi_{\rm min}^{\rm NOON} = \frac{\sqrt{(1-l)^{-n}+1}}{\sqrt{2}n}$ \cite{PhysRevA.75.053805}. The corresponding behavior is shown in Fig.~\ref{fig:Standard MZI with coherent losses}. As expected, the phase sensitivity of the NOON state deteriorates rapidly with increasing loss, highlighting the well-known fragility of entangled states to photon losses \cite{Kacprowicz2010}. The MZI with coherent-vacuum input is comparatively more robust, although its phase sensitivity also worsens significantly as the photon losses increase. In contrast, the HA protocol for $N=1$ and $N=3$ as shown in Fig.~\ref{fig:HA-enhanced MZI losses with leff}  exhibits considerably greater robustness against losses.
\begin{figure}[H]
\centering
\includegraphics[scale = 0.675]{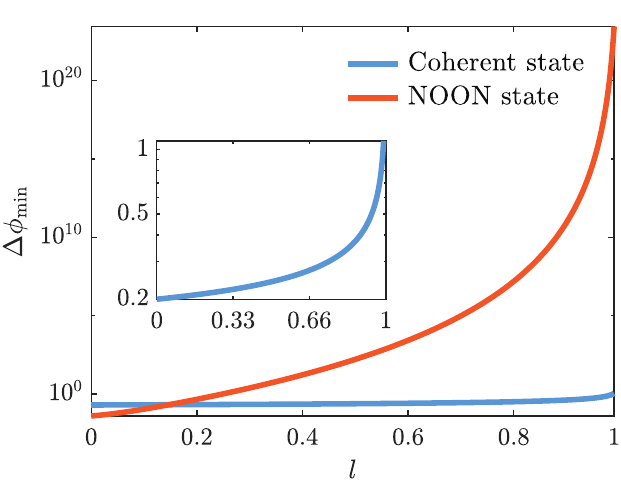}
\caption{Minimum phase uncertainty $\Delta\phi_{\rm min}$ as a function of the loss parameter $l$ for conventional MZI with coherent-vacuum $\ket{\psi} = \ket{\alpha}_{a_{0}} \otimes \ket{0}_{b_{0}}$ input with $|\alpha|^2 =25$ (blue) and for NOON state $\ket{\psi_{\rm NOON}}=\frac{1}{\sqrt{2}}(\ket{0}_{a} \otimes \ket{n}_{b}+\ket{n}_{a} \otimes\ket{0}_{b})$ with $n=25$ (orange). The MZI with coherent vacuum exhibits a gradual degradation with increasing loss and diverges only in the limit of complete photon loss, $l\rightarrow 1$. In contrast, the NOON state is highly sensitive to losses, with its phase sensitivity deteriorating rapidly even for moderate values of $l$. The inset shows a magnified view of the coherent-state result.}
\label{fig:Standard MZI with coherent losses}
\end{figure}

\end{document}